\documentclass[12pt]{elsarticle}

\usepackage[left = 1in, right = 1in]{geometry}
\usepackage{amssymb}
\usepackage{amsmath}
\usepackage{relsize}
\usepackage{nicefrac,xfrac}
\usepackage{gensymb}
\usepackage{graphicx}
\usepackage{multirow}
\usepackage{subcaption}
\usepackage{algorithm}
\usepackage{algpseudocode}
\usepackage{adjustbox}
\usepackage{nomencl}

\makenomenclature
\algnewcommand{\LineComment}[1]{\State \(\triangleright\) #1}

\begin{document}

\begin{frontmatter}

\title{Iteration-Free Cooperative Distributed MPC through Multiparametric Programming}

\author[inst1]{Radhe S. T. Saini}

\affiliation[inst1]{organization={Department of Chemical Engineering},
            addressline={Indian Institute of Technology Gandhinagar}, 
            city={Palaj, Gandhinagar},
            postcode={382355}, 
            state={Gujarat},
            country={India}}

\author[inst2]{Parth R. Brahmbhatt}

\author[inst2]{Styliani Avraamidou\corref{cor1}}
% \email{avraamidou@wisc.edu}
\author[inst1]{Hari S. Ganesh\corref{cor1}}

\affiliation[inst2]{organization={Department of Chemical and Biological Engineering},%Department and Organization
            addressline={University of Wisconsin-Madison}, 
            city={Madison},
            postcode={WI 53706}, 
            country={U.S.A.}}

\cortext[cor1]{Corresponding authors. \\ \textit{E-mail addresses}:  avraamidou@wisc.edu and hariganesh@iitgn.ac.in}

\begin{abstract}

Cooperative Distributed Model Predictive Control (DiMPC) architecture employs local MPC controllers to control different subsystems, exchanging information with each other through an iterative procedure to enhance overall control performance compared to the decentralized architecture. However, this method can result in high communication between the controllers and computational costs. In this work, the amount of information exchanged and the computational costs of DiMPC are reduced significantly by developing novel iteration-free solution algorithms based on multiparametric (mp) programming. These algorithms replace the iterative procedure with simultaneous solutions of explicit mpDiMPC control law functions. The reduced communication among local controllers decreases system latency, which is crucial for real-time control applications. The effectiveness of the proposed iteration-free mpDiMPC algorithms is demonstrated through comprehensive numerical simulations involving groups of coupled linear subsystems, which are interconnected through their inputs and a cooperative plant-wide cost function.  

\end{abstract}


\end{frontmatter}

\newpage

\begin{nomenclature}
    \nomenclature{$\textbf{x}$}{State vector}
    \nomenclature{$\textbf{u}$}{Input vector}
    \nomenclature{$A$}{System matrix}
    \nomenclature{$B$}{Input matrix}
    \nomenclature{$k$}{Sample time instant}
    \nomenclature{$q$}{Sample time step-related index}
    \nomenclature{$n_x$}{Dimension of state vector}
    \nomenclature{$n_{x,i}$}{Dimension of state vector of the $i^{th}$ subsystem}
    \nomenclature{$n_u$}{Dimension of the input vector}
    \nomenclature{$n_{u,j}$}{Dimension of the input vector from the $j^{th}$ local model predictive controller (or of the subsystem)}
    \nomenclature{$J$}{Objective function of model predictive control optimal control problem}
    \nomenclature{$N_p$}{Prediction horizon}
    \nomenclature{$Q$}{State weight matrix}
    \nomenclature{$R$}{Input weight matrix}
    \nomenclature{$P$}{Terminal state weight matrix}
    \nomenclature{$\chi$}{State constraint set}
    \nomenclature{$\Omega$}{Terminal state constraint set}
    \nomenclature{$\mathcal{U}$}{Input constraint set}
    \nomenclature{$\textbf{U}$}{Control input decision vector}
    \nomenclature{$\textbf{U}^*$}{Optimal control input decision vector}
    \nomenclature{$j|k$}{$j$ steps ahead prediction based on known information at time instant $k$}
    \nomenclature{$\theta$}{Vector of unknown (or uncertain) parameters}
    \nomenclature{$H$}{Quadratic term matrix in the multiparametric optimal control problem objective}
    \nomenclature{$H_t$}{Linear term matrix in the multiparametric optimal control problem objective}
    \nomenclature{$c$}{Constant term vector in the multiparametric optimal control problem objective}
    \nomenclature{$G$}{Constraint matrix in the multiparametric optimal control problem}
    \nomenclature{$b$}{Constraint vector in the multiparametric optimal control problem}
    \nomenclature{$F$}{Parameter matrix in the multiparametric optimal control problem}
    \nomenclature{$n_{c}$}{Number of constraints in the multiparametric optimal control problem}
    \nomenclature{$M$}{Number of subsystems (or local model predictive controllers)}
    \nomenclature{$i, j$}{Subsystem indices}
    \nomenclature{$\rho_i$}{Weight for subsystem $i$ in the distributed model predictive control optimal control problem objective}
    \nomenclature{$p$}{Intermediate iteration index}
    \nomenclature{$\textbf{w}$}{Weight for convex combination in the iterative process}
    \nomenclature{$\epsilon$}{Error tolerance for convergence}
    \nomenclature{$p_{max}$}{Maximum number of intermediate iterations}
    \nomenclature{$CR^{v}$}{$v^{th}$ Critical region}
    \nomenclature{$n_{CR}$}{Number of critical regions}
    \nomenclature{$\mathbb{R}$}{Set of real numbers}
    \nomenclature{$\Phi^{v}$}{Matrix that define the $v^{th}$ critical region}
    \nomenclature{$\phi^{v}$}{Vector that define the $v^{th}$ critical region}
    \nomenclature{$\mathbf{0}$}{Zero vector}
    \nomenclature{$r$}{Intermediate iteration index}
    \nomenclature{$\bar{\textbf{U}}$}{Converged control input vector after intermediate iterations}
    \nomenclature{$\bar{\theta}$}{Vector of unknown parameters which considers only states}
    \nomenclature{$\bar{\textbf{V}}_i$}{Control input vector of all local model predictive controllers except the $i^{th}$ controller}
\end{nomenclature}

\printnomenclature
\newpage

\section{Introduction}

Model Predictive Control (MPC) is a widely-implemented optimal control technique that emerged from chemical plants, oil refineries, and process industries in the late 1980s~\cite{cutler1980dynamic, qin1997overview}. Since its inception, many innovative and improved versions of MPC have been developed and deployed for various industrial process control systems~\cite{qin1997overview, morari1999model}. MPC solves an optimal control problem (OCP) by utilizing predictive models to calculate a sequence of control inputs over a specified future time horizon at each sample time step in a receding horizon manner. The OCP typically aims to minimize a tracking objective over a future time horizon, subject to input, output, equipment, and operational  constraints~\cite{rawlings2017model, maciejowski2002predictive}. Despite its ability to handle constraints and calculate optimized inputs, control of large-scale systems presents significant difficulties for MPC~\cite{trodden2017distributed}. As the complexity of the plant structure increases, fetching feedback signals and applying control signals to farther areas of the plant increases the communication burden. Moreover, the centralized MPC (CMPC) architecture, in which the control inputs are calculated by solving the OCP using the plantwide objective function at each sample time step, can result in large computational overhead. This can impede the ability of the controller to perform real-time calculations within one sample time period determined by process dynamics and operating conditions. Alternatively, decentralized MPC (DeMPC) architecture, with its ability to decouple the entire plant into a set of subsystems, reduces complexities. However, such decoupling impedes DeMPC from capturing subsystem interactions, thereby resulting in sub-optimal control performance and instabilities~\cite{cui2002performance, siljak2011decentralized}.

Distributed MPC (DiMPC) architecture serves as a compromise between the centralized and the decentralized control architectures, coupling the relative advantages of both. DiMPC combines the decoupled architecture of DeMPC with the ability to capture subsystem interactions of CMPC. DiMPC architecture can be further classified into non-cooperative and cooperative variants~\cite{negenborn2014distributed, christofides2013distributed, avraamidou2017multi, richards2007robust, venkat2007distributed}. In non-cooperative DiMPC, each local controller optimizes its control inputs based on local information and the information received from other subsystems without considering the effects of its control actions on the overall system. In contrast, in cooperative DiMPC, each local controller optimizes its control inputs by considering not only local information but also the effects of its control actions on the overall system performance, as quantified by a global objective function ~\cite{farina2012distributed,  ferramosca2013cooperative,  stewart2010cooperative}. Nevertheless, in both non-cooperative and cooperative DiMPCs, the local control inputs are exchanged among the local controllers and the optimal solution is found by performing intermediate iterations or solving the OCP several times to ideally converge to the solution at each sample time step. However, only cooperative DiMPC results converge to the centralized case (the best case) with the increase in the number of intermediate iterations~\cite{stewart2010cooperative, ferramosca2013cooperative}. Hence, cooperative DiMPC will be the focus of this work, and from now on, `DiMPC' will refer to the cooperative case. 

Recent advancements in DiMPC have focused on improving theoretical foundations and algorithm efficiency. Wang and Yang~\cite{wang2022improved} developed an improved iterative solution method to accelerate convergence while reducing the computational burden. Darivianakis et al.~\cite{darivianakis2019distributed} introduced a DiMPC approach that ensures robust constraint satisfaction for systems with coupled uncertainties. Recent applications of DiMPC span diverse domains, showcasing its versatility and effectiveness in complex systems. In energy systems, DiMPC has been applied to optimize the operation of integrated energy hubs, enhancing synergy and grid response through decomposition and cooperation strategies \cite{wu2023distributed}. For vehicle platooning, DiMPC algorithms have been developed to ensure string stability and consensus among vehicles, improving safety and efficiency in autonomous driving scenarios \cite{feng2023distributed}. In the field of building energy management, DiMPC has been employed to control distributed energy resources in residential buildings, considering forecast uncertainties to optimize energy consumption and costs \cite{langner2024model}. Additionally, DiMPC has shown promise in wind farm control, coordinating individual turbines to optimize overall power output and grid integration \cite{lu2024advances}, and in irrigation systems, where it addresses challenges of hydraulic coupling and computational demands \cite{zhu2023distributed}. These applications demonstrate the broad applicability and ongoing advancements in DiMPC across various engineering disciplines.

A major drawback of DiMPC, however, is the increased communication load and computational costs resulting from the need to solve multiple OCPs and exchange information among local controllers within each sample time period. Hence, in practice, intermediate iterations are terminated early or before converging to the solution according to termination criteria, resulting in sub-optimal control performance.  Note that, in DiMPC, the overall system is decomposed into smaller coupled subsystems and each local controller solves an OCP with a plantwide objective function and local variables as decision variables or control inputs. Although the OCPs of local controllers can be solved in parallel, the computational costs pertaining to the large number of intermediate iterations and the communication load to exchange local control inputs through a shared communication network remain high, usually at prohibitive levels.

Two different approaches to overcome the drawback of DiMPC have been reported in the literature. The first approach involves improving the rate of convergence of the intermediate iterations~\cite{wang2022improved}. Fewer intermediate iterations to converge to the solution reduces the computational cost. This can also improve the control performance as at several instants where the intermediate iterations had to be terminated early, the optimal, converged solutions can be obtained within one sample time period. The second approach involves reducing the time of calculation of each intermediate iteration. This is achieved by reformulating and solving the DiMPC OCP using multiparametric (mp) programming~\cite{pistikopoulos2002line, bemporad2002explicit} by also considering the information from other local controllers as unknown parameters, thereby generating the control law offline as affine functions of system parameters. This offline-generated control law is expressed as a set of critical regions, each corresponding to a distinct region in the parameter space. The offline generated expressions reduce online calculations (at each intermediate iteration) to critical region search followed by algebraic function evaluation, thereby reducing the computation time of the DiMPC controller~\cite{saini2023noncooperative, diangelakis2016decentralized, kirubakaran2014distributed}. This iterative mpDiMPC controller will be referred to as I-mpDiMPC from now onward.

Although I-mpDiMPC helps reduce the computational costs, the method may not significantly reduce the communication burden among the local controllers. Reducing the communication burden significantly would require the distributed control solution scheme to be iteration-free. A few studies have been performed by restricting the local controllers to exchange information only once per sample time step, thus making the control scheme iteration-free. Camponogara~et~al.~\cite{camponogara2002distributed} studied a DiMPC controller that communicates the previous time step control inputs of local controllers among themselves at each sample time step. Similar iteration-free approaches were also studied in which each local controller sends its future control inputs to the other local controllers at each sampling time~\cite{chen2016improved, farina2011distributed}. In all such studies, the local controllers use shared information delayed by one sample time step, thereby not guaranteeing global solution and stability in DiMPC and I-mpDiMPC. While DiMPC and I-mpDiMPC can attain global-like conditions when allowed to iterate until convergence is attained, the use of shared information delayed by one sample time step may not result in a similar DiMPC architecture as when no time delay is applied.

The novelty of this work lies in developing state-of-the-art, iteration-free mpDiMPC algorithms to significantly improve the computational cost and communication load compared to the DiMPC algorithm. The novel algorithms aim to reduce the number of information exchanges between the local controllers, thereby offering efficiency gains and potential robustness benefits by mitigating the latency issues from communication overhead. The idea is to generate mp programming functions for all the local controllers offline and then solve them simultaneously, instead of following an iterative procedure, to obtain the optimal control inputs at each sample time instant. By exchanging information only once per sample period, the proposed algorithms minimize communication delays and are less susceptible to the effects of lost or corrupted data during communication.

The remainder of this paper is organized as follows. In Section~\ref{sec: Background}, currently existing DiMPC control architecture and mp programming is explained. In Section~\ref{sec: ProposedAppraches}, three different versions of the developed iteration-free mpDiMPC algorithms are presented. A set of case studies using random plants is defined in Section~\ref{sec: CaseStudy}. The results and discussion are presented in Section~\ref{sec: results} and conclusions are drawn in Section~\ref{sec: conclusions}.

\section{MPC and iterative DiMPC overview} \label{sec: Background}

This section provides an overview of the formulation and solution methods for MPC and DiMPC and their mp programming versions, mpMPC and I-mpDiMPC, respectively. 

Consider the plant dynamics of a system given by: 
\begin{equation} \label{eq: MPC_LTI_Model}
    \begin{split}
    \begin{aligned}
    \textbf{x}(k + 1) & = A\textbf{x}(k) + B\textbf{u}(k) \\
    \end{aligned}
    \end{split}
\end{equation}

\noindent where, $\textbf{x} \in \mathbb{R}^{n_x}$ is the state vector, $\textbf{u} \in \mathbb{R}^{n_u}$ is the input vector, $A \in \mathbb{R}^{n_x \times n_x}$ is the system matrix, $B \in \mathbb{R}^{n_x \times n_u}$ is the input matrix, and \textit{k} is the sample time instant.

\subsection{MPC} \label{sec: MPC}
In standard or online MPC, a single OCP whose decision variables represent the future control inputs of the plant is solved at each sample time step. The OCP with the objective function given by:
\begin{equation}
\begin{aligned}
    J(\textbf{x}(k), \textbf{U}(k)) = \Bigg\{ & \sum_{j = 0}^{N_p-1} \Bigg[\frac{1}{2}\textbf{x}^T(j|k)Q\textbf{x}(j|k)
     + \frac{1}{2}\textbf{u}^T(j|k)R\textbf{u}(j|k)\Bigg] \\  
    & + \frac{1}{2}\textbf{x}^T(N_p|k)P\textbf{x}(N_p|k) \Bigg\} 
\end{aligned}
\end{equation}

\noindent is generally formulated as: 
\begin{equation}
\begin{aligned}
   \min_{\textbf{U}(k)} &  J(\textbf{x}(k), \textbf{U}(k))\\
    \text{s.t.} & ~ \textbf{x}(l+1|k) = A\textbf{x}(l|k) + B\textbf{u}(l|k) , \forall ~ l \in \mathbb{I}_{0:N_p-1} \\
    & \textbf{x}(l|k) \in \chi \ \forall \ l \in \mathbb{I}_{1:N_p} \\
    & \textbf{x}(N_p|k) \in \Omega \\
    & \textbf{u}(l|k) \in \mathcal{U} \ \forall \ l \in \mathbb{I}_{0:N_p-1} \\
    & \textbf{x}(0|k) = \textbf{x}(k)
\end{aligned}
\label{eq: CMPCFormulation}
\end{equation}

\noindent where, $\chi$, $\Omega$, and $\mathcal{U}$ are the bounding sets for state vector, terminal state vector, and control input vector, respectively, $\textbf{U}(k) = [\textbf{u}(0|k), \textbf{u}(1|k), ..., \textbf{u}(N_p-1|k)]$ is the control input decision variable vector, $j|k$ is the $j$ steps ahead prediction from the current time $k$, $l$ is a time step related index, $Q$ and $R$ are the weight matrices for states and control inputs, respectively, $P$ is the terminal state weight matrix, and $N_p$ is the prediction horizon. Solving equation~\eqref{eq: CMPCFormulation} results in the optimal decision vector $\textbf{U}^*(k)$, among which, the first control input $\textbf{u}^*(k) \ (= \textbf{u}(0|k))$ is implemented in the system. The aforementioned procedure is repeated and the OCP is solved again at the next sample time instant $k+1$ using the new state information vector $\textbf{x}(k+1)$.

\subsection{mpMPC}

The standard MPC described in Section~\ref{sec: MPC} is also referred to as online MPC because the OCP is solved at each sample time step $k$ to obtain the optimized control inputs. The same results can be obtained by determining the
explicit expressions of the control law offline through the mp programming approach, reducing online calculations to point location and function evaluation. Using the state-space model of the system given by eq.~\eqref{eq: MPC_LTI_Model}, the future states can be expressed as a function of the current state $\textbf{x}(k)$ and future control inputs as:
\begin{equation}
    \textbf{x}(l+1|k) = A^{l+1}\textbf{x}(k) + \mathlarger{\sum}_{q=0}^{l} A^qB\textbf{u}(l-q|k)
\label{eq: futureStatesUpdateEquation}
\end{equation}

\noindent where $q$ is the sample time step-related index. In the standard MPC OCP given by eq.~\eqref{eq: CMPCFormulation}, the state vector $\textbf{x}(k)$ is usually the only uncertainty (unknown parameter) which becomes known online at each sample time instant. However, variables such as the previous control input information, measured output information, varying set points, etc., could also be considered as unknown parameters if they are made available only online during controller deployment. Defining the set of the unknown parametric variables as $\theta$ vector ($\theta = \textbf{x}(k)$) and substituting eq.~\eqref{eq: futureStatesUpdateEquation} in eq.~\eqref{eq: CMPCFormulation} to reformulate the OCP into the following:

\begin{equation}
\begin{aligned}
    \min_{\textbf{U}} \frac{1}{2}\textbf{U}^TH\textbf{U} + (\theta ^TH_{t} + c)\textbf{U} \\
    G\textbf{U} \leq b + F\theta
\end{aligned}
\label{eq: multiparametricProblem}
\end{equation}

\noindent where, $H \in \mathbb{R}^{n_u \times n_u}$, $H_{t} \in \mathbb{R}^{n_x \times n_u}$, $c \in \mathbb{R}^{n_u}$, $G \in \mathbb{R}^{n_c \times n_u}$, $b \in \mathbb{R}^{n_c}$,$F \in \mathbb{R}^{n_c \times n_x}$ , and $n_{c}$ is the number of constrains.
The problem can be solved multiparametrically to express the optimal control inputs as explicit affine functions of vector $\theta$, which remain both valid and optimal in a specific polyhedral space or critical region. The explicit (multiparametric) solution to eq.~\eqref{eq: multiparametricProblem} can be expressed mathematically as:

\begin{equation}
 \begin{aligned}
     \textbf{U}^* = f^v(\theta) ~~ if ~~ CR^v : \Phi^v\theta \leq \phi^v \\  
     \forall ~~ v = 1, 2, 3, ..., n_{CR}
 \end{aligned}
 \label{eq: mpMPCSol}
 \end{equation}

 \noindent where $f$ represents the affine function defined for each critical region $v$ with $\Phi \ \text{and} \ \phi$ as the corresponding inequality matrices and $n_{CR}$ represents the total number of critical regions. During online deployment, at a given time instant $k$, with the availability of $\theta$, the critical region $CR^{v*}$  which satisfies $\Phi^{v*}\theta \leq \phi^{v*}$ is identified and the corresponding function $f^{v*}(\theta$) is evaluated to obtain the optimal control input vector $\textbf{U}^*(k)$. The first element in $\textbf{U}^*(k)$, i.e., $\textbf{u}^*(k) \ (= \textbf{u}(0|k))$, is implemented in the system. The aforementioned procedure is repeated and region search followed by function evaluation is performed again at the next time instant $k+1$, using the new information of the vector $\theta$.

\subsection{DiMPC} \label{sec: DiMPC}

In DiMPC, the plant is decomposed into a number of subsystems, wherein the local subsystem model also considers the interaction dynamics among the subsystems. An array of local controllers calculates the local control inputs. The local controllers exchange information with each other and iteratively solve a plantwide objective function to determine the control inputs. At each sample time step, the control inputs of the local controllers are calculated through an iterative procedure by performing intermediate iterations. At each intermediate iteration, every local controller calculates its (intermediate) control inputs based on the local information and the latest (intermediate) control input information received from other local controllers. At the next intermediate iteration step, the local controller typically uses the latest (intermediate) control inputs of the other local controllers to calculate the (intermediate) control inputs. The iterative procedure is repeated until a pre-defined termination criterion is reached. Upon termination, the (intermediate) control input of the latest intermediate iteration is implemented in the plant as the control input at the particular sample time step. At the next sample time step, the control inputs implemented in the plant are treated as the latest (intermediate) control inputs to start the iterative procedure to determine the control inputs of all the local controllers.   

Consider the dynamics of the $i^{th}$ subsystem among a network of $M$ subsystems given by:
\begin{equation}
    \textbf{x}_i(k+1) = A_i\textbf{x}_i(k) + \sum_{j=1}^M B_{i,j}\textbf{u}_j(k)
\end{equation}

\noindent where, $\textbf{x}_i \in \mathbb{R}^{n_{x,i}}$, $ A_i \in \mathbb{R}^{n_{x,i} \times n_{x,i}}$, $B_{i,j} \in \mathbb{R}^{n_{x,i} \times n_{u,j}}$, and $\textbf{u}_j \in \mathbb{R}^{n_{u,j}}$. The overall system can be assembled as:

\begin{equation}
    \textbf{x}(k+1) = A\textbf{x}(k) + \sum_{j=1}^M B_{j}\textbf{u}_j(k)
    \label{eq: overall_system_model}
\end{equation}

\noindent where, $\textbf{x}(k) = [\textbf{x}_1(k), \textbf{x}_2(k),..., \textbf{x}_M(k)] \in \mathbb{R}^{n_x}$, $n_x = \sum_{i = 1}^M n_{x,i}$, $A = diag(A_1, A_2, ..., A_M)$, and $B_j = [B_{1,j}^T, B_{2,j}^T, ..., B_{M,j}^T]^T$, where, $diag()$ represents a block-diagonal matrix. Assume a local cost function of the $i^{th}$ local controller given by:
\begin{equation}
\begin{aligned}
    J_i(\textbf{x}_i(k), \textbf{U}_{i}(k)) = \sum_{j = 0}^{N_p-1} \Bigg[\frac{1}{2}\textbf{x}_i^T(j|k)Q_i\textbf{x}_i(j|k) &\\ + \frac{1}{2}\textbf{u}_i^T(j|k)R_i\textbf{u}_i(j|k)\Bigg]  \\
     + \frac{1}{2}\textbf{x}_i^T(N_p|k)P_i\textbf{x}_i(N_p|k)
\end{aligned}
\end{equation}

\noindent where, $\textbf{U}_{i}(k) = [\textbf{u}_i(0|k), $ $\textbf{u}_i(1|k), ..., \textbf{u}_i(N_p-1|k)]$ is the predicted input trajectory. The plantwide objective function can be accumulated as:
\begin{equation}
    J(\textbf{x}(k), \textbf{U}(k)) = \sum_{i=1}^M \rho_iJ_i(\textbf{x}_i(k), \textbf{U}_{i}(k))
\end{equation}

\noindent where, $\rho_i > 0$ is the relative weight and $\textbf{U}(k) = [\textbf{U}_1(k), \textbf{U}_2(k), ..., \textbf{U}_M(k)]$.

Let $p$ be any positive integer representing an intermediate iteration. At any time step \textit{k}, the OCP solved by every $i^{th}$ controller at intermediate iteration $p$ is:
\begin{equation}
\begin{aligned}
   \mathcal{F}_i^{(p)} =  \min_{\textbf{U}_i^{(p)}(k)} & ~ J(\textbf{x}(k), \textbf{U}(k)) \\
    \text{s.t.} & ~ \textbf{x}(l+1|k) = A\textbf{x}(l|k) + \sum_{j=1}^M B_{j}\textbf{u}_j(l|k),\\ & \forall l \in \mathbb{I}_{0:N_p-1} \\
    & \textbf{x}_t(l|k) \in \chi_t \ \forall \ l \in \mathbb{I}_{1:N_p}, \forall \ t \in \mathbb{I}_{1:M}\\
    & \textbf{x}_t(N_p|k) \in \Omega_t \ \forall \ t \in \mathbb{I}_{1:M} \\
    & \textbf{u}_i (l|k) \in \mathcal{U}_i \ \forall \ l \in \mathbb{I}_{0:N_p-1} \\
    & \textbf{x}(0|k) = \textbf{x}(k), \\
    & \textbf{U}_j(k) = \textbf{U}_j^{(p-1)}(k), \ \forall  \ j \in \mathbb{I}_{1:M \backslash i}
\end{aligned}
\label{eq: i^th_DMPCProblem}
\end{equation}

\noindent where, $\mathbb{I}_{1:M \backslash i}$ represents the set of all $M$ local controllers except the $i^{th}$ controller, i.e., \{$1, 2, ..., i-1, i+1, ...,M$\}. During the optimization process, the control inputs of other local controllers $\textbf{U}_j(k)$, $\ \forall \ j \in \mathbb{I}_{1:M \backslash i}$ remain unchanged, as ensured by the last equality condition in the aforementioned OCP. Once the optimal control inputs by all $M$ controllers at $p^{th}$ intermediate iteration is obtained, the next iterate is calculated as a convex combination of the current and previous iterate as follows:
\begin{equation}
    \textbf{U}^{(p)}(k) = \textbf{w}\textbf{U}^{(p-1)}(k) + (1-\textbf{w})\textbf{U}^{(p)}(k)
\end{equation}

\noindent The values of weight $\textbf{w}$ dictate the convergence abilities of the iterative process. An adaptive algorithm to accelerate the convergence process as described in ~\cite{wegstein1958accelerating} is used in this study. Essentially, the method tries to manipulate $\textbf{w}$ at each iterative step using the results of the previous time step. Note that $\textbf{w} = [ w_1, w_2, ..., w_i, ..., w_m ]$ contains individual weights for each subsystem $i \in {1, 2, ..., M}$, ensuring that for each subsystem, $w_i$ and its complement $(1 - w_i)$ sum to 1, where $w_i \in[0,1]$. At the $1^{st}$ intermediate iteration, a warm start is used as the initial condition as defined below:
\begin{equation}
\begin{aligned}
    \textbf{U}^{(0)}(k) = \Bigg\{ & \textbf{u}_1(1|k-1), \textbf{u}_1(2|k-1), ...,\\ & \textbf{u}_1(N_p-1|k-1), \textbf{0}_{1 \times n_{u,1}}, \\
    & \textbf{u}_2(1|k-1), \textbf{u}_2(2|k-1), ...,\\ & \textbf{u}_2(N_p-1|k-1), \textbf{0}_{1 \times n_{u,2}}, ..., \\
    & \textbf{u}_M(1|k-1), \textbf{u}_M(2|k-1), ..., \\ & \textbf{u}_M(N_p-1|k-1), \textbf{0}_{1 \times n_{u,M}}, \Bigg\}
\end{aligned}
\label{eq: warmstart}
\end{equation}

\noindent The warm start is assembled using the optimal solution found at the previous sample time step $(k-1)$. At any $k^{th}$ sample time step, the DiMPC algorithm incorporated with Wegenstein's approach is described in Algorithm~\ref{alg:DiMPC}. At any $k^{th}$ sample time step, a warm start vector $\bar{\textbf{U}}^{(1)}$ is assembled according to eq.~\eqref{eq: warmstart}. At each intermediate iteration, the OCP of each local controller defined in eq.~\eqref{eq: i^th_DMPCProblem} is solved. The weight vector $\textbf{w}$ is modified based on optimal results from current and previous intermediate iterations. The optimal control inputs at each intermediate iteration $p$ are saturated between control limits. Once the convergence criteria are satisfied, the intermediate iterations are terminated and the control input obtained at the latest $p^{th}$ iteration is implemented in the system.

\begin{algorithm}
    \caption{DiMPC (or I-mpDiMPC) algorithm applying the Wegenstein's approach for accelerated convergence}\label{alg:DiMPC}
    \begin{algorithmic}
    \Require $\textbf{U}(k-1), \textbf{x}(k), \epsilon > 0, p_{max} \geq 0$
    \State $\bar{\textbf{U}}^{(1)} = \textbf{U}^{0}(k)$ \Comment{Assemble using $\textbf{U}(k-1)$ according to eq.~\eqref{eq: warmstart}}
    \State $ r = 1$
        \For{$p = 1 \ \text{to} \ p_{max}$}
            \For{$i = 1 \ \text{to} \ M$}
                \LineComment{For DiMPC, solve eq.~\eqref{eq: i^th_DMPCProblem} with arguments $\textbf{x}(k)$, and the current control input information 
                $\bar{\textbf{U}}^{(p)}$ received from other subsystems}
                \LineComment{For I-mpDiMPC, use pre-computed solutions given by eq.~\eqref{eq: M1_DiMPCSol}}
                \State $ \textbf{U}^{(p+1)} = \arg(\mathcal{F}_i^{(p)})$
            \EndFor
            \If{$r = 1$}
                \State $\bar{\textbf{U}}^{(p-1)} = \bar{\textbf{U}}^{(p)}$
                \State $\bar{\textbf{U}}^{(p)} = \textbf{U}^{(p+1)}$
                \State $r = 2$
            \Else
                \LineComment{Element wise vector manipulations}
                \State $\textbf{a} = \sfrac{(\textbf{U}^{(p+1)} - \textbf{U}^{(p)})}{(\bar{\textbf{U}}^{(p)} - \bar{\textbf{U}}^{(p-1)})}$
                \State $\textbf{w} =  \sfrac{\textbf{a}}{(\textbf{a} - 1)}$
                \State $\textbf{w} = \max(\min(\textbf{w}, w_{max}), w_{min})$
                \State $\bar{\textbf{U}}^{(p+1)} = \textbf{w}\bar{\textbf{U}}^{(p)} + (1-\textbf{w})\textbf{U}^{(p+1)}$
                \State $\bar{\textbf{U}}^{(p-1)} = \bar{\textbf{U}}^{(p)}$
                \State $\bar{\textbf{U}}^{(p)} = \bar{\textbf{U}}^{(p+1)}$
            \EndIf
            \State $\textbf{U}^{(p)} = \textbf{U}^{(p+1)}$
            \State $\bar{\textbf{U}}^{(p)} = \max(\min(\bar{\textbf{U}}^{(p)}, \textbf{U}_{max}), \textbf{U}_{min})$
            \State Transmit $\bar{\textbf{U}}^{(p)}$ to all interconnected systems
            \If{$\text{all}(| \bar{\textbf{U}}^{(p)} - \bar{\textbf{U}}^{(p-1)}| < \epsilon)$}
                \State \textbf{break out of for loop}
            \EndIf
            \State $p = p+1$
        \EndFor
        \State $\textbf{U}^*(k) = \bar{\textbf{U}}^{(p)}$ \Comment{Optimal or best possible control solution at time step \textit{k}}
    \end{algorithmic}
\end{algorithm}

% \subsection{Multiparametric DiMPC formulation  (mpDiMPC)} \label{sec: mpDiMPC}

\subsection{I-mpDiMPC} \label{sec: mpDiMPC}

Similar to the solution method described in Section~\ref{sec: DiMPC}, a multiparametric version of DiMPC can be developed as follows. Using the overall system model (see eq.~\eqref{eq: overall_system_model}), all future states can be expressed as a function of current state $\textbf{x}(k)$ and future control actions as:
\begin{equation}
    \textbf{x}(l+1|k) = A^{l+1}\textbf{x}(k) + \mathlarger{\sum}_{q=0}^{l} A^q\sum_{j=1}^MB_j\textbf{u}_j(l-q|k)
\label{eq: DiMPC_futureStatesUpdateEquation}
\end{equation}

\noindent For every $i^{th}$ local controller, control actions of all other local controllers $\textbf{U}_j \ \forall \ j \in \mathbb{I}_{1:M \backslash i}$ are treated as unknown parameters along with the state vector $\textbf{x}(k)$. Grouping them in a single unknown parametric vector as $\theta_i = [ \textbf{x}(k), \textbf{U}_1, \textbf{U}_2, ..., \textbf{U}_{i-1}, \textbf{U}_{i+1}, ..., \textbf{U}_M]$, substituting eq.~\eqref{eq: DiMPC_futureStatesUpdateEquation} in eq.~\eqref{eq: i^th_DMPCProblem} and rearranging results in:
\begin{equation}
\begin{aligned}
    \mathcal{F}_i^{(p)}(\theta_i) = \min_{\textbf{U}_i} \frac{1}{2}\textbf{U}_i^TH_i\textbf{U}_i + (\theta_i^TH_{t,i} + c_i)\textbf{U}_i \\
    G_i\textbf{U}_i \leq b_i + F_i\theta_i
\end{aligned}
\label{eq: DiMPC_multiparametricProblem}
\end{equation}

\noindent The above problem can be solved multiparametrically to determine the optimal control inputs as explicit affine functions of vector $\theta_i$, which remain both valid and optimal in a specific polyhedral space, termed as critical region. The multiparametric solution to eq.~\eqref{eq: DiMPC_multiparametricProblem} is:
\begin{equation}
 \begin{aligned}
     \textbf{U}_i^{(p)} = f_i^v(\theta_i) ~~ if ~~ CR_i^v : \Phi_i^v\theta_i \leq \phi_i^v \\  %  W_i^v(\theta_i) + w_i^v,
     \forall ~~ v = 1, 2, 3, ..., n_{CR,i}
 \end{aligned}
 \label{eq: M1_DiMPCSol}
 \end{equation}

 \noindent where, $f_i$ is an affine function evaluated for each critical region of every $i^{th}$ OCP. The algorithm for this approach is similar to that of Algorithm~\ref{alg:DiMPC} for DiMPC or I-mpDiMPC, except that the optimal control input for every $i^{th}$ controller is calculated from the offline-determined explicit solutions.

\section{Novel iteration-free mpDiMPC algorithms} \label{sec: ProposedAppraches}

To make the aforementioned I-mpDiMPC algorithm iteration-free, it is further developed through the procedure of simultaneous solution of the offline-generated control law expressions of the local controllers during online deployment. The simultaneous solution procedure replaces the critical region search procedure performed at every intermediate iteration of the I-mpDiMPC at each sample time instant, making the online deployment iteration-free. Three different iteration-free mpDiMPC algorithms, namely, IF-mpDiMPC, IF-mpDiMPC-V1.5, and IF-mpDiMPC-V2, are developed and their computational and control performance is studied. The iteration-free methods enable the mpDiMPC controllers to calculate the control inputs with minimal communication in the distributed control network, improving the resilience of the overall system to issues arising from excessive communication.

% \subsection{Concurrent multiparametric DiMPC (CmpDiMPC)}

\subsection{IF-mpDiMPC}

In the first method, i.e., IF-mpDiMPC, the optimal control inputs at each sample time step are determined by simultaneously solving the explicitly generated affine functions of the control law for all the local controllers in the network. The developed algorithm is explained in detail in this subsection. 

For a local IF-mpDiMPC controller $i$, among other unknown parameters, $\theta_i$ will also contain control input information of all other local controllers. Splitting the parametric vector $\theta_i$ as:

\begin{equation}
    \theta_i = \big[\underbrace{\textbf{x}(k)}_\text{$\Bar{\theta}$}, \ \underbrace{\textbf{U}_1, \textbf{U}_2, ..., \textbf{U}_{i-1}, \textbf{U}_{i+1}, ... \textbf{U}_M}_\text{$\Bar{\textbf{V}}_i$} \ \big]
\end{equation}

\noindent where, $\Bar{\textbf{V}}_i$ is a subset of $\textbf{U}$ that includes the control inputs for all local controllers except the $i^{th}$ one, i.e., $[\textbf{U}_1, ..., \textbf{U}_{i-1}, \textbf{U}_{i+1}, ..., \textbf{U}_M]$, and $\Bar{\theta}$ is the remaining unknown parameters. In this study, $\Bar{\theta} = \textbf{x}(k)$. Substituting $\theta_i$ in eq.~\eqref{eq: M1_DiMPCSol} yields the expression of the control law given by:

 \begin{equation}
 \begin{aligned}
     \textbf{U}_i = f_i^v(\Bar{\theta}, \Bar{\textbf{V}}_i) ~~ if ~~ CR_i^v : \Phi_{i,1}^v\Bar{\theta} + \Phi_{i,2}^v\Bar{\textbf{V}}_i \leq \phi_i^v \\
     \forall ~~ v = 1, 2, 3, ..., n_{CR,i}
 \end{aligned}
 \label{eq: M1_DiMPCSol2}
 \end{equation}

\noindent where, $\Phi_{i}^v = \begin{bmatrix}
    \Phi_{i,1}^v & \textbf{0} \\ \textbf{0}  & \Phi_{i,2}^v
\end{bmatrix}$ is a block-diagonal matrix. Similarly, multiparametric solutions for all the $M$ local controllers can be determined and the explicit expressions of the control law can be assembled as given below:

 \begin{equation}
 \begin{aligned}
 \begin{matrix}
     \textbf{U}_1 = f_1^v(\Bar{\theta}, \Bar{\textbf{V}}_1) ~~ if ~~ CR_1^v : \Phi_{1,1}^v\Bar{\theta}_1 + \Phi_{1,2}^v\Bar{\textbf{V}}_1 \leq \phi_1^v
         \\ \forall ~~ v = 1, 2, 3, ..., n_{CR,1} \\
    \textbf{U}_2 = f_2^v(\Bar{\theta}, \Bar{\textbf{V}}_2) ~~ if ~~ CR_2^v : \Phi_{2,1}^v\Bar{\theta}_2 + \Phi_{2,2}^v\Bar{\textbf{V}}_2 \leq \phi_2^v \\ \forall ~~ v = 1, 2, 3, ..., n_{CR,2} \\
     . \\
     . \\
     . \\
     \textbf{U}_M = f_M^v(\Bar{\theta}, \Bar{\textbf{V}}_M) ~~ if ~~ CR_M^v : \Phi_{M,1}^v\Bar{\theta}_M + \Phi_{i,2}^v\Bar{\textbf{V}}_M \leq \phi_M^v \\ \forall ~~ v = 1, 2, 3, ..., n_{CR,M}
 \end{matrix}
 \end{aligned}
 \label{eq: M1_DiMPCSol3}
 \end{equation}

  % where $\Bar{\textbf{V}}_i$ is a subset of $\textbf{U}$ that includes the control inputs for all subsystems except the $i^{th}$ one, i.e., $[\textbf{U}_1, ..., \textbf{U}_{i-1}, \textbf{U}_{i+1}, ..., \textbf{U}_M]$.

\noindent By considering $\textbf{U}$ as $\Bar{\textbf{V}}_i$ and $\textbf{U}_i$, through appropriate matrix manipulations, the aforementioned equation can be expressed in terms of $\textbf{U}$ as:
\begin{equation}
 \begin{aligned}
 \begin{matrix}
     \textbf{U} = g_1^v(\Bar{\theta}) ~~ if ~~ CR_1^v : \Phi_{1,1}^v\Bar{\theta}_1 + \Phi_{1,2}^v\Bar{\textbf{V}}_1 \leq \phi_1^v \
         \\ \forall ~~ v = 1, 2, 3, ..., n_{CR,1} \\
    \textbf{U} = g_2^v(\Bar{\theta}) ~~ if ~~ CR_2^v : \Phi_{2,1}^v\Bar{\theta}_2 + \Phi_{2,2}^v\Bar{\textbf{V}}_2 \leq \phi_2^v \
     \\ \forall ~~ v = 1, 2, 3, ..., n_{CR,2} \\
     . \\
     . \\
     . \\
     \textbf{U} = g_M^v(\Bar{\theta}) ~~ if ~~ CR_M^v : \Phi_{M,1}^v\Bar{\theta}_M + \Phi_{i,2}^v\Bar{\textbf{V}}_M \leq \phi_M^v \
     \\ \forall ~~ v = 1, 2, 3, ..., n_{CR,M}
 \end{matrix}
 \end{aligned}
 \label{eq: M1_DiMPCSol4}
 \end{equation}

\noindent The affine functions of the control law expressions result in formulating $N_p\sum_{i=1}^M n_{u,i}$ equations to evaluate same number of unknowns and therefore can be solved simultaneously (instead of the iterative procedure) to yield the final optimal control vector $\textbf{U}^*(k)$. Note that the unknown parametric vector $\Bar{\theta}$ consists of a known state vector. The pseudo-code to implement IF-mpDiMPC is provided in Algorithm~\ref{alg: mpCoDiMPC}.

\begin{algorithm}
\caption{IF-mpDiMPC algorithm}\label{alg: mpCoDiMPC}
\begin{algorithmic}
\Require $\textbf{x}(k)$
\State $\Bar{\theta} = \textbf{x}(k)$
\For{$L_1 = 1 \ to \ n_{CR,1}$}  \Comment{$L_1, ..., L_M$ are indices of critical regions for subsystems $1$ through $M$}
\For{$L_2 = 1 \ to \ n_{CR,2}$}
\State .
                    \For{$L_i = 1 \ to \ n_{CR,i}$}
                        \State .
                            \For{$L_M = 1 \ to \ n_{CR,M}$}
                                \State Simultaneously solve the equations,\\$\textbf{U} = g_1^{L_1}(\Bar{\theta}), \textbf{U} = g_2^{L_2}(\Bar{\theta}), ...,$ $\textbf{U} = g_i^{L_i}(\Bar{\theta}), ..., \textbf{U} = g_M^{L_M}(\Bar{\theta})$
                            \EndFor \Comment{$M^{th} \ for \ loop$}
                        \State .
                    \EndFor \Comment{$i^{th} \ for \ loop$}
                \State .
            \EndFor \Comment{$2^{nd} \ for \ loop$}
        \EndFor \Comment{$1^{st} \ for \ loop$}
        \State $\textbf{U}^*(k) = \textbf{U}$ \Comment{Optimal control solution at time step \textit{k}}
    \end{algorithmic}
\end{algorithm}

Solving the affine functions simultaneously for all possible combinations of critical regions space results in obtaining the final optimal control vector. When in conflict with multiple feasible solutions, the one that results in the least value of the objective in the OCP (see eq.~\eqref{eq: DiMPC_multiparametricProblem}) is selected. An important consideration for the practical implementation of the proposed IF-mpDiMPC approach is the impact of the number of critical regions on the computational complexity of the simultaneous solution step. Specifically, the simultaneous equations need to be solved $n_{CR,1}\times n_{CR,2} \times...\times n_{CR,M}$ times. The active critical region combination is determined conclusively by solving the simultaneous equations for all possible combinations, which can be computationally expensive. Hence, the IF-mpDiMPC-V1.5 and IF-mpDiMPC-V2 algorithms are developed to further improve the computational speed of the IF-mpDiMPC algorithm. The approach is to reduce the possible combinations of critical regions or the number of times the simultaneous solutions need to be performed on a set of critical regions online at a given sample time step.

\subsection {IF-mpDiMPC-V1.5}

In the second method, i.e., IF-mpDiMPC-V1.5, to reduce the computation time for IF-mpDiMPC, the number of critical regions that have to be considered for simultaneous solution is reduced by performing feasibility checks on the inequality constraints of the critical regions. The developed algorithm is explained in detail in this subsection. 

Consider the explicit solution of the local controller $1$ among all $M$ controllers, i.e., the first equation in eq.~\eqref{eq: M1_DiMPCSol4}:
 \begin{equation}
 \begin{matrix}
 \textbf{U} = g_1^v(\Bar{\theta}) ~~ if ~~ CR_1^v : \Phi_{1,1}^v\Bar{\theta}_1 + \Phi_{1,2}^v\Bar{\textbf{V}}_1 \leq \phi_1^v \\ \forall ~~ v = 1, 2, 3, ..., n_{CR,1} \\
 \end{matrix}
 \end{equation}

\noindent Representing the expressions of the critical region inequalities as:
 \begin{equation}
 \begin{matrix}
 \textbf{U} = g_1^v(\Bar{\theta}) ~~ if ~~ CR_1^v : \Phi_{1,2}^v\Bar{\textbf{V}}_1 \leq \phi_1^v - \Phi_{1,1}^v\Bar{\theta}_1\
       \\  \forall ~~ v = 1, 2, 3, ..., n_{CR,1} \\
\end{matrix}
 \end{equation}

\noindent the initial formulation of inequality matrices, $\Phi_{1,1}^v, \ \Phi_{1,2}^v, \ \text{and} ~ \phi_1^v$ are performed with a feasibility check during the multiparametric solution development. Now, a subset of available information $\Bar{\theta_1}$ can be used to determine feasibility such that the inequality constraints corresponding to a critical region are satisfied. Consequently, this feasibility check would ensure that the optimal solution might lie in this critical region. On the other hand, if the $\Bar{\theta_1}$ value violates the inequality constraints corresponding to a critical region, then this would suggest that the optimal solution will not lie in the critical region. This initial check can help reduce the possible number of critical regions (to $n_{CR,1}^* (\leq n_{CR,1})$) that have to be accounted for in simultaneous solution for IF-mpDiMPC to obtain the optimal control inputs. 

The aforementioned procedure is then followed for all the multiparametric, explicitly generated expressions of the control law for all $M$ local controllers. For each local controller, the subset of the feasible critical regions $n_{CR,i}^* \leq n_{CR,i}, \forall i \in 1,2,3,... M$ are identified. Consequently, the number of times the simultaneous equations should be solved for the overall system to reduced to the following:

\begin{equation}
\begin{matrix}
   n_{CR,1}^*\times n_{CR,2}^* \times...\times n_{CR,M}^* \leq \\ n_{CR,1}\times n_{CR,2} \times...\times n_{CR,M} 
\end{matrix}  
\end{equation}

\noindent The check for feasibility for the set of inequalities in the $v^{th}$ critical region for each $i^{th}$ controller can be found by solving the following linear programming (LP) problem with a dummy objective:
\begin{equation}
\begin{aligned}
    h_{1,v}^* = & \min_{\Bar{\textbf{V}}_1} 0 \\
    & \Phi_{1,2}^v\Bar{\textbf{V}}_1 \leq \phi_1^v - \Phi_{1,1}^v\Bar{\theta}_1\
\end{aligned}\label{eq:mCmpCpDiMPC-LP}
\end{equation}

\noindent The psuedocode to implement IF-mpDiMPC-V1.5 is provided in Algorithm~\ref{alg: mpImCoDiMPC}.

\begin{algorithm}
    \caption{IF-mpDiMPC-V1.5 algorithm}\label{alg: mpImCoDiMPC}
    \begin{algorithmic}
        \Require $\textbf{U}(k-1), \textbf{x}(k)$
        \State $\Bar{\theta} = \textbf{x}(k)$
        \LineComment{Accumulate $n_{CR,i}^*$ for all i local controllers}
        \For{$i = 1\ to\ M$} \Comment{Outer loop to generate feasible sets for M local controllers}
        \State $N_{CR,i} = \{ \}$ \Comment{An empty set to store indices of all feasible critical regions}
            \For{$Ii = 1 \ to \ n_{CR,i}$}
                \State Formulate LP problem as in eq.~\eqref{eq:mCmpCpDiMPC-LP}
                \If{$h_{1,Ii}^*$ is non empty}
                    \State Append $Ii$ to $N_{CR,i}$
                \EndIf
            \EndFor
        \EndFor
        \For{$L_1 \ \text{in} \ N_{CR,1}$} \Comment{$L_1, ..., L_M$ are indices of critical regions for subsystems $1$ through $M$}
            \For{$L_2 \ \text{in} \ N_{CR,2}$}
                \State .
                    \For{$L_i \ \text{in} \ N_{CR,i}$}
                        \State .
                            \For{$L_M \ \text{in} \ N_{CR,M}$}
                                \State Simultaneously solve the equations,\\ $\textbf{U}^* = g_1^{L_1}(\Bar{\theta}), \textbf{U}^* = g_2^{L_2}(\Bar{\theta}),...,$ $\textbf{U}^* = g_i^{L_i}(\Bar{\theta}), ..., \textbf{U}^* = g_M^{L_M}(\Bar{\theta})$
                            \EndFor \Comment{$M^{th} \ for \ loop$}
                        \State .
                    \EndFor \Comment{$i^{th} \ for \ loop$}
                \State .
            \EndFor \Comment{$2^{nd} \ for \ loop$}
        \EndFor \Comment{$1^{st} \ for \ loop$}
        \State $\textbf{U}^*(k) = \textbf{U}^*$ \Comment{Optimal control solution at time step \textit{k}}
    \end{algorithmic}
\end{algorithm}

\subsection{IF-mpDiMPC-V2}

In the third method, i.e., IF-mpDiMPC-V2, the computational costs of IF-mpDiMPC-V1.5 are further reduced by exploring only the nearest neighbors of the optimal critical region solution of the previous sample time step. The developed algorithm is explained in detail in this subsection. 

In practice, the plant states change gradually. Hence, gradual movements in control inputs would be sufficient to steer the plant to its desired operating condition. Therefore, the control input solution will likely lie in and around the optimal critical region identified at the previous sample time step. 

Assume the control input solution at $(k-1)^{th}$ sample time step is $\textbf{U}^*(k-1) = [\textbf{U}_1(k-1), \textbf{U}_2(k-1), ..., \textbf{U}_M(k-1)]$. For the $i^{th}$ local controller at sample time step $k$, an unknown parametric vector is formed using current state information $\textbf{x}(k)$ given by:
\begin{equation}
\begin{matrix}
    \theta_i = \big[\textbf{x}(k), \textbf{U}_1(k-1), \textbf{U}_2(k-1), ..., \textbf{U}_{i-1}(k-1), \\ \textbf{U}_{i+1}(k-1), .., \textbf{U}_M(k-1) \big]    
\end{matrix}
\end{equation}

\noindent Based on the unknown parametric vector $\theta_i$, a search for the feasible and valid critical region is conducted and the optimal critical region for $i^{th}$ local controller is identified. In the next sample time step $k+1$, a set consisting of this critical region and its nearest neighbors is formed and considered for the simultaneous solution for the control inputs calculation. Note that this set would be smaller than the set considered for the simultaneous solution, not only in IF-mpDiMPC but also in IF-mpDiMPC-V1.5. The pseudo-code for the IF-mpDiMPC-V2 algorithm is provided in Algorithm~\ref{alg: mpNNDiMPC}.

\begin{algorithm}
    \caption{IF-mpDiMPC-V2 algorithm}\label{alg: mpNNDiMPC}
    \begin{algorithmic}
        \Require $\textbf{U}(k-1), \textbf{x}(k)$
        \State \textbf{do}
            \For{$i = 1\ to\ M$} \Comment{Outer loop to generate feasible sets for M local controllers}
                \State \ \ \ $N_{CR,i} = \{ \}$ \Comment{An empty set to store indices of all feasible critical regions}
                \State \ \ \ Form Unknown parametric vector $\theta_i = \big[\textbf{x}(k),\textbf{U}_1(k-1),...,\textbf{U}_{i-1}(k-1),\textbf{U}_{i+1}(k-1), ..., \textbf{U}_M(k-1) \big]$
                \State \ \ \ Perform Point Location algorithm and find the optimal critical region $v_1$ using explict solution of $1^{st}$ local controller and the unknown parametric vector $\theta_1$.
                \State \ \ \ Form the feasible set $N_{CR,i}$ using $v_i$ and its nearest boundaries. Thus, $N_{CR,i} = \{v_i, \text{boundaries of }v_i \}$
            \EndFor
        \State \textbf{end do}
        \For{$L_1 \ \text{in} \ N_{CR,1}$}  \Comment{$L_1, ..., L_M$ are indices of critical regions for subsystems $1$ through $M$}
            \For{$L_2 \ \text{in} \ N_{CR,2}$}
                \State .
                    \For{$L_i \ \text{in} \ N_{CR,i}$}
                        \State .
                            \For{$L_M \ \text{in} \ N_{CR,M}$}
                                \State Simultaneously solve the equations, \\ $\textbf{U} = g_1^{L_1}(\Bar{\theta}), \textbf{U} = g_2^{L_2}(\Bar{\theta}), ...,$ $\textbf{U} = g_i^{L_i}(\Bar{\theta}), ..., \textbf{U} = g_M^{L_M}(\Bar{\theta})$
                            \EndFor \Comment{$M^{th} \ for \ loop$}
                        \State .
                    \EndFor \Comment{$i^{th} \ for \ loop$}
                \State .
            \EndFor \Comment{$2^{nd} \ for \ loop$}
        \EndFor \Comment{$1^{st} \ for \ loop$}
        \If{$\textbf{U}$ not found}\\
        Revert to iterative I-mpDiMPC approach (Algorithm 1) to obtain $\textbf{U}$
        \EndIf
        \State $\textbf{U}^*(k) = \textbf{U}$ \Comment{Optimal control solution at time step \textit{k}}
    \end{algorithmic}
\end{algorithm}
If the optimal control input solution is not found in the critical region or its nearest neighbors from the previous time step, the algorithm reverts to the iterative I-mpDiMPC approach (see Section~\ref{sec: mpDiMPC}). This ensures convergence to the optimal solution or early termination at the latest sub-optimal solution if the sample time is completed, bounding the worst-case execution time of IF-mpDiMPC-V2 by that of I-mpDiMPC. Future research could explore alternative fallback strategies for large-scale systems.

The mp programming explicit expressions of the control law for all the local controllers are the same for both I-mpDiMPC and the developed IF-mpDiMPC algorithms. However, the IF-mpDiMPC algorithms reformulate the expressions into simultaneous equations. Note that the reformulation does not alter the underlying solution space or the optimal solutions obtained through multiparametric programming. Hence, the stability properties of the iterative algorithms are also applicable to the developed IF-mpDiMPC algorithms. Therefore, there is no need to prove the stability of the developed solution methods or algorithms. 

\section{Case studies} \label{sec: CaseStudy}

To study the performance of the developed iteration-free algorithms, they are implemented on systems consisting of $M \in [2,3,4,5]$ subsystems. Each subsystem is assumed to be composed of a plant with two states and one input $(n_{x,i} = 2, n_{u,i} = 1, \forall i \in 1:M)$. Each element of state matrix $A_i$ and input matrix $B_{i,j}$ is randomly chosen from the interval $[-1, 1]$. The coupling between subsystems is inherent in the structure of the randomly generated plants, specifically in the $B_{i,j}$ matrices. Also, the states are assumed to be upper bounded by a random number from the interval $[10, 100]$ and lower bounded by a number chosen from the interval $[-100, -10]$. Similarly, the upper bound of inputs is chosen from the interval $[1, 5]$ and the lower bound is chosen from $[-5, -1]$. The prediction horizon length is assumed to equal the control horizon length for all local controllers and is set to 3 in this work. $100$ random controllable plants are created for each case study of $M \in [2,3,4,5]$ with dynamic matrices and initial conditions chosen randomly within the aforementioned bounds. For example, a random plant with $2$ subsystems is:
\begin{equation}
    \begin{aligned}
        \textbf{x}_1(k+1) = A_1\textbf{x}_1(k) + B_{1,1}u_1(k) + B_{1,2}u_2(k) \\
        \textbf{x}_2(k+1) = A_2\textbf{x}_2(k) + B_{2,1}u_1(k) + B_{2,2}u_2(k) 
    \end{aligned}
\end{equation}

\noindent where, $\textbf{x}_1 \in \mathbb{R}^2$, $\textbf{x}_2 \in \mathbb{R}^2$,
    $A_1 = \begin{bmatrix}
    0.1645  &   0.7399  \\
    0.0815  &   -0.4704 
    \end{bmatrix}$, $B_{1,1} = \begin{bmatrix}
    -0.3639 \\  -0.7616
    \end{bmatrix}$, $B_{1,2} = \begin{bmatrix}
    0.8797  \\  0.2911 
    \end{bmatrix}$, $A_2 = \begin{bmatrix}
    -0.0411  &   0.0894  \\
    0.2786  &   0.2946 
    \end{bmatrix}$, $B_{2,1} = \begin{bmatrix}
    0.0878  \\  0.4421 
    \end{bmatrix}$, and $B_{2,2} = \begin{bmatrix}
    0.0450  \\ 0.9874 
    \end{bmatrix}$. Further, the state and input dynamic vectors are bounded as follows:

    \begin{equation}
        \begin{aligned}
            \begin{bmatrix}
                -63.5878    \\   -59.6464
            \end{bmatrix} \leq  \textbf{x}_1 \leq \begin{bmatrix}
                    29.6809 \\ 19.5218
            \end{bmatrix} \\
            \begin{bmatrix}
                -67.0765    \\  -31.2846
            \end{bmatrix} \leq  \textbf{x}_2 \leq \begin{bmatrix}
                19.8728 \\  15.7232
            \end{bmatrix} \\
        -1.2686 \leq u_1 \leq 3.5116 \\
        -1.1090 \leq u_2 \leq 4.0879
        \end{aligned}
    \end{equation}

\section{Results and Discussion} \label{sec: results}

\begin{figure}[h!]
    \centering
    \includegraphics[width=0.7\columnwidth]{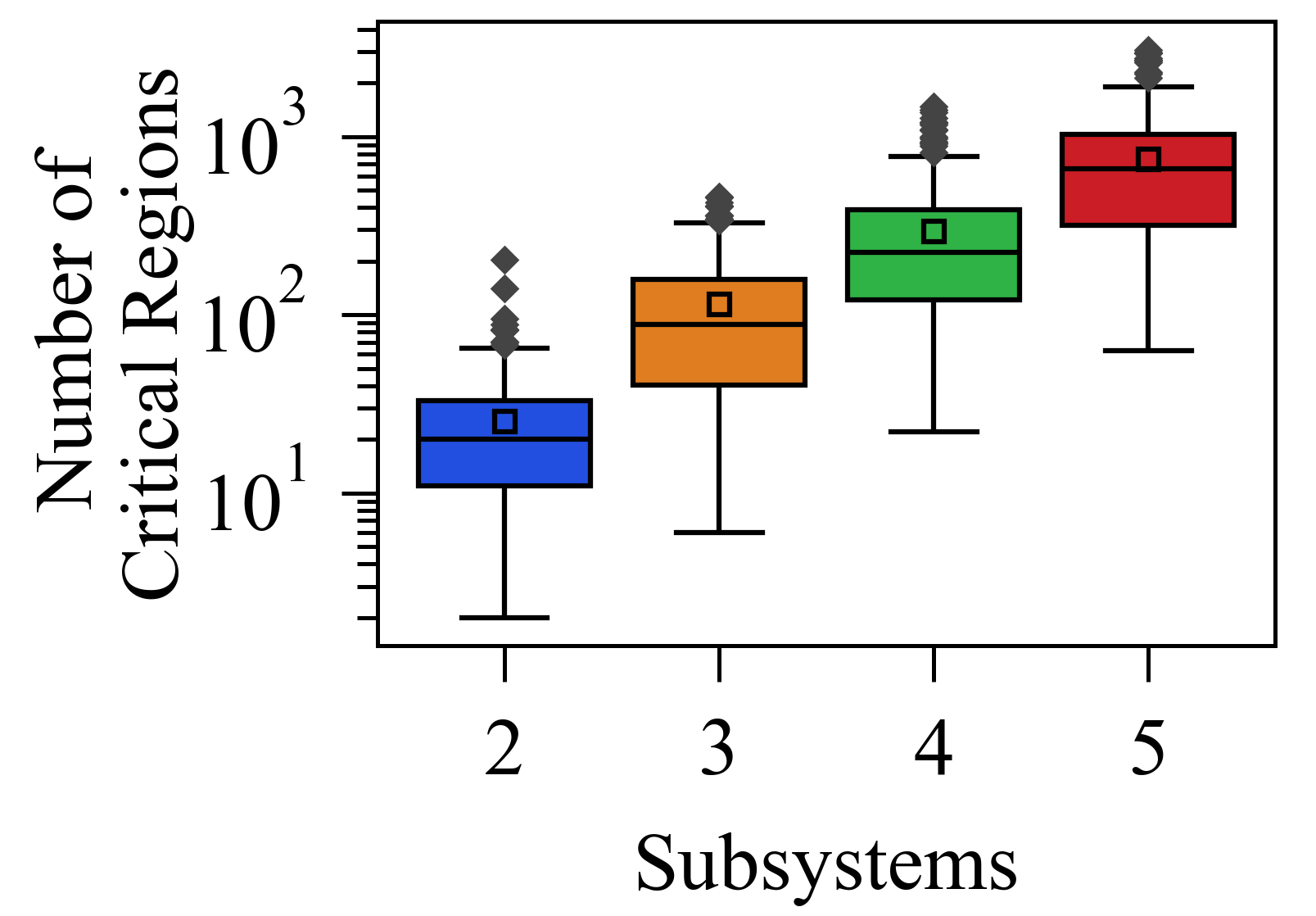}
    \caption{The number of critical regions for different number of subsystems on a log scale. Note that the number of critical regions remains the same for all the mp programming DiMPC solution approaches, which differ only in the calculation method of optimal control inputs.}
    \label{fig:critical_regions}
\end{figure}

In this section, the numerical simulation results of the random plants under the developed IF-mpDiMPC controllers are compared in terms of trade-offs between computational efficiency and communication load with the standard DiMPC~\cite{stewart2010cooperative} and I-mpDiMPC controllers. All simulations are performed on a desktop computer with Intel(R) Core(TM) i7-10700 CPU @ 2.90GHz processor, 16 GB RAM, and 512 GB SSD. The quadratic programming (QP) problems in our DiMPC algorithm are solved using the IBM ILOG CPLEX \cite{cplex2009v12} solver. For the explicit multiparametric quadratic programming (mpQP) solutions required in I-mpDiMPC and all versions of IF-mpDiMPC, the POP Toolbox \cite{oberdieck2016pop} is employed. Also, default conditions have been retained while using the CPLEX QP solver and POP toolbox while simulating all case studies. Initially, random plants as described in Section~\ref{sec: CaseStudy} are generated. Subsequently, the explicit solutions of each local I-mpDiMPC controller in all random plants are calculated. The boundaries of each critical region are also determined offline and stored (for use during online deployment). Figure~\ref{fig:critical_regions} shows the distribution of the total number of critical regions of all the local controllers combined for the different subsystem cases. Note that the number of critical regions remains the same for all mp programming approaches,  I-mpDiMPC, IF-mpDiMPC, IF-mpDiMPC-V1.5, and  IF-mpDiMPC-V2, as only the calculation of the optimal control input solutions from the critical regions differ in these algorithms. It can be observed that the number of critical regions increases exponentially with the number of systems.

\begin{figure}
  \centering
        \begin{subfigure}[b]{0.8\columnwidth}
            \includegraphics[width=\linewidth]{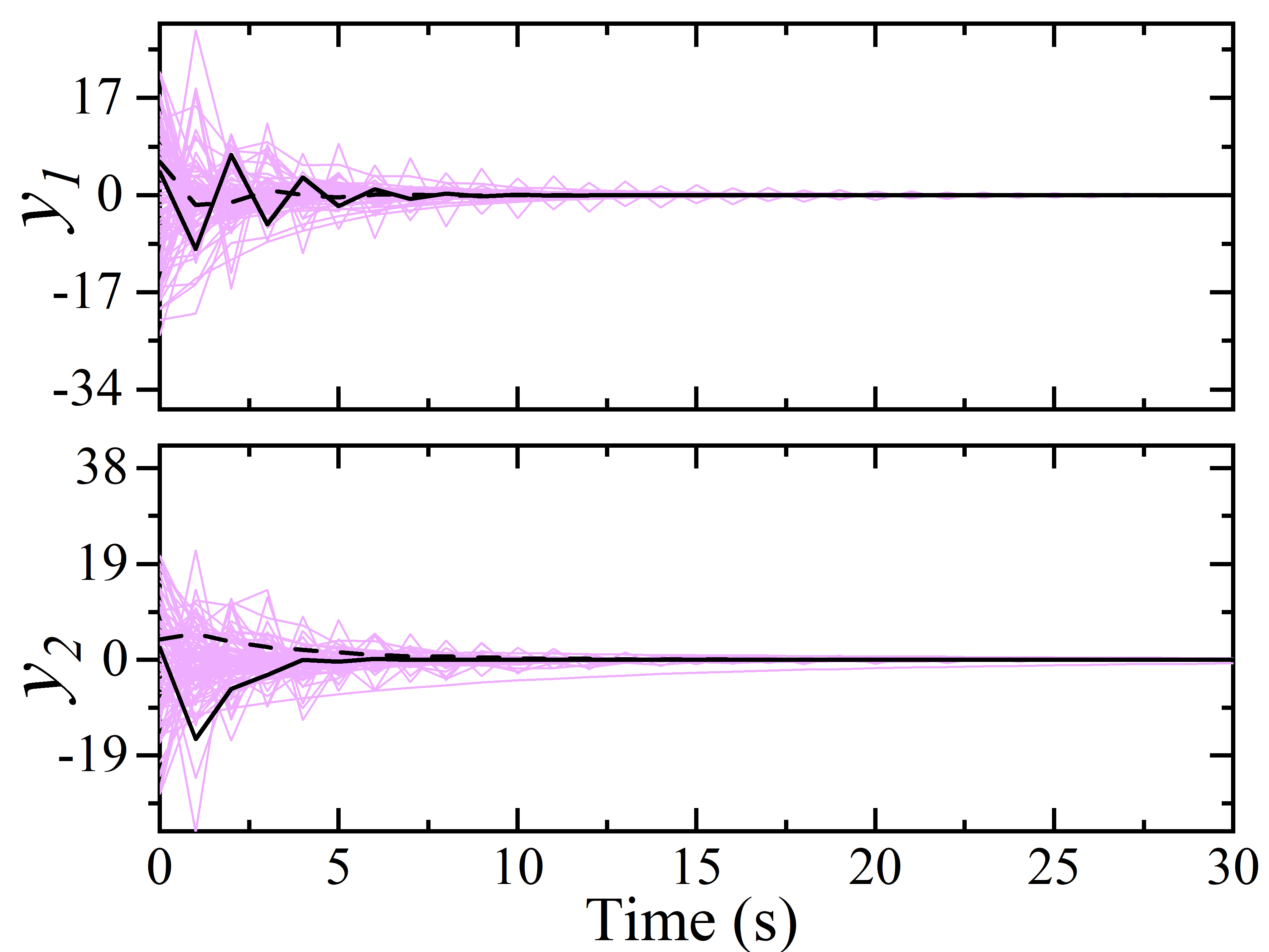}
            \caption{}
            \label{fig:2SubsOP}
        \end{subfigure}
        \begin{subfigure}[b]{0.8\columnwidth}
            \includegraphics[width=\linewidth]{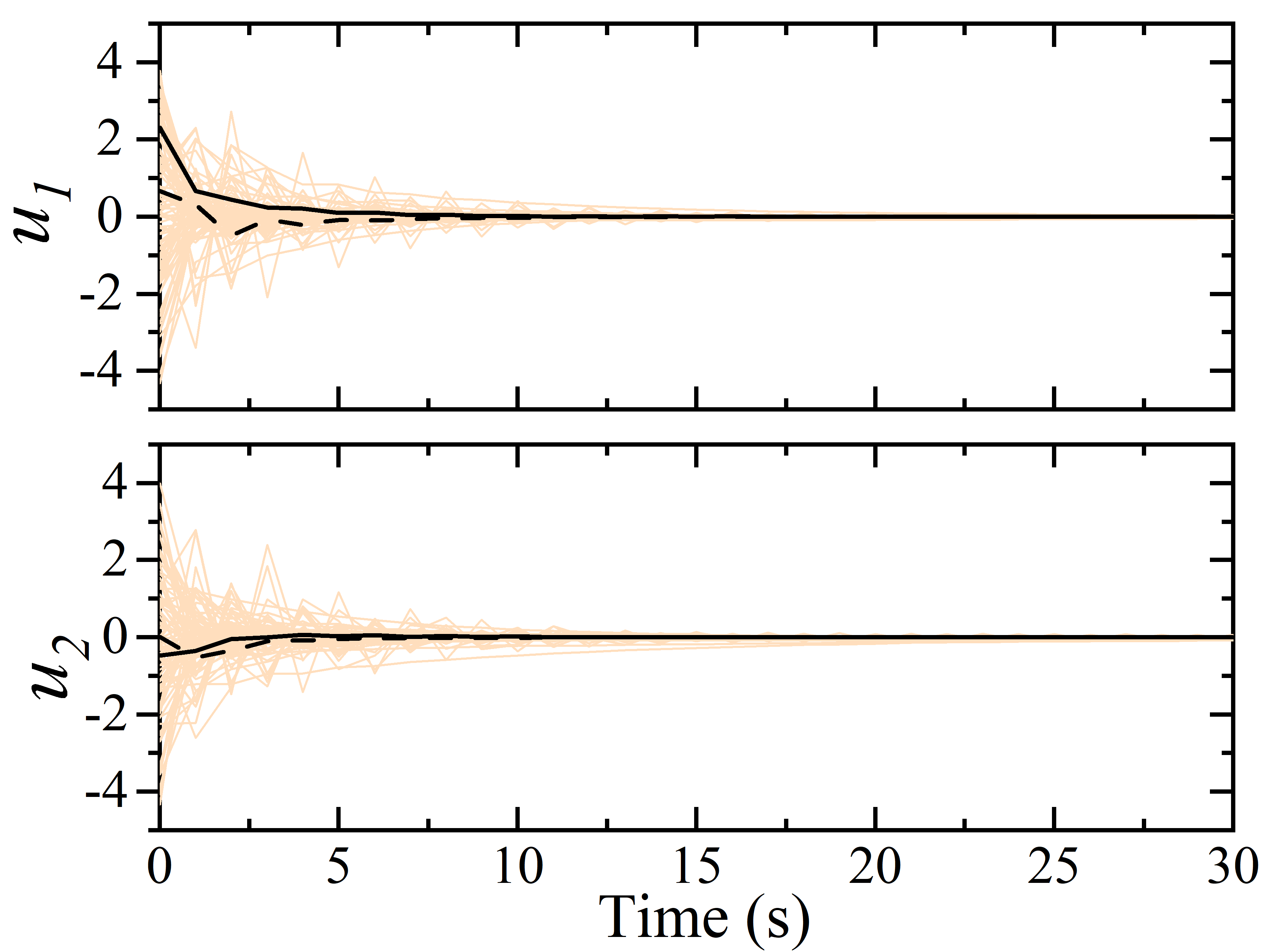}
            \caption{}
            \label{fig:2SubsIP}
        \end{subfigure}\\
  \caption{Controller performance for 2 subsystems. (a) Outputs of each subsystem. (b) Applied inputs to achieve the given outputs. The solid and dashed lines represent two random experimental runs in the output and input plots for visualization purposes.}
  \label{fig:2SubsPerformances}
\end{figure}

The random plants are simulated for a time duration of $100$ s with random initial conditions (kept the same for all solution algorithms for comparison purposes) at time $0$ s. For the concise representation of the results, only a single output value, the sum of all states for each subsystem, is shown. For example, in case of $M = 2$ subsystems (Figure~\ref{fig:2SubsPerformances}), wherein each subsystem controller has 2 states $(\textbf{x}_1 \in \mathbb{R}^2, \text{and} ~ \textbf{x}_2 \in \mathbb{R}^2)$, the outputs $y_1$ and $y_2$ are the element-wise summations of the state vectors $\textbf{x}_1$ and $\textbf{x}_2$, respectively. While this approach simplifies the presentation of results across the randomly generated plant models, in physical systems, summing different state variables may not always have a meaningful interpretation. The focus here is on demonstrating the overall dynamic behavior and convergence of the subsystems under different control strategies.

To ensure that the control performance results of DiMPC and I-mpDiMPC match with those of the developed IF-mpDiMPC algorithms, a low value of $10^{-8}$ is considered for the error tolerance $\epsilon$ (see Algorithm~\ref{alg:DiMPC}). Also, the maximum number of allowable intermediate iterations $p_{max} = 100$. The closed-loop simulation results for the 2-subsystem case (see Fig.~\ref{fig:2SubsPerformances}) demonstrate that all controllers achieve centralized-like performance due to the low value of $\epsilon$. The closed loop simulation results representing the control performance of the 3 - 5 subsystem cases also demonstrate centralized-like performance and can be found the supplementary material. The controllers successfully steer random initial conditions to equilibrium within approximately 30 s, despite the randomly selected plant dynamics. While iterative approaches like DiMPC and I-mpDiMPC require more iterations and incur higher communication and computational costs to achieve plantwide optimality, the proposed IF-mpDiMPC method achieves similar performance efficiently, reducing both communication and computational overhead by simultaneously solving the explicitly generated control law expressions of the different local controllers.

\begin{figure*}[ht!]
  \centering
  \includegraphics[width=1\textwidth]{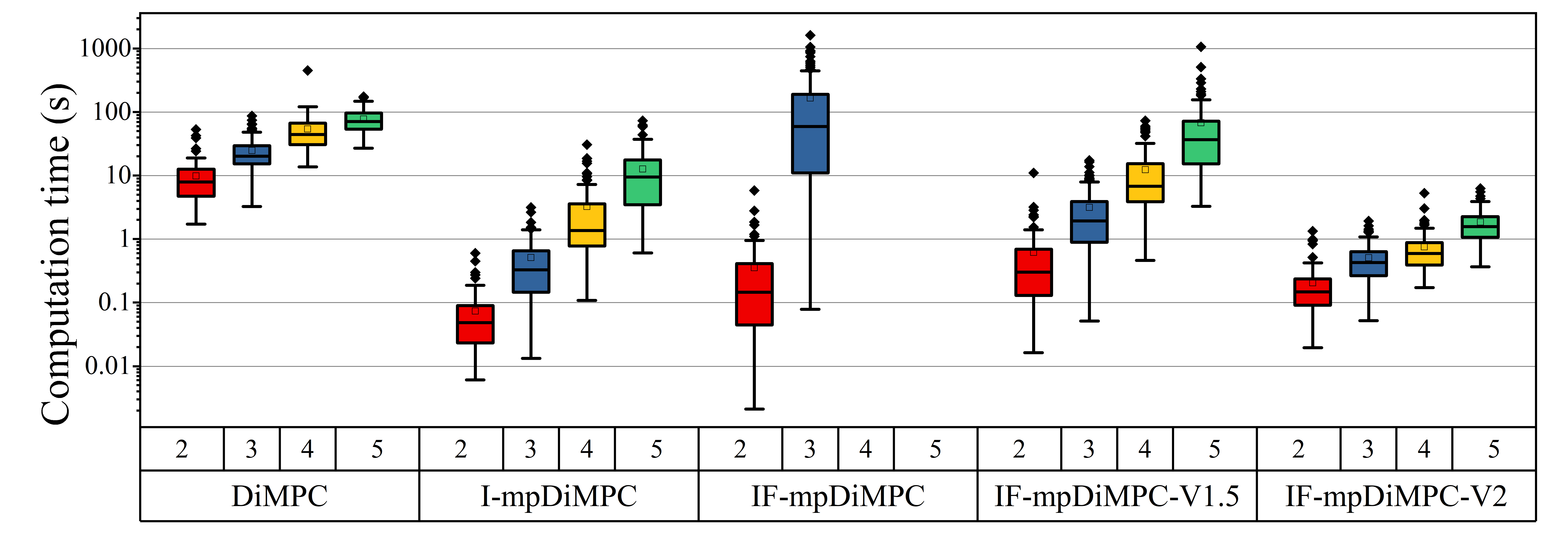}
  \caption{Computation times in log scale for the different DiMPC control architectures.}
  \label{fig:CompTimesI1}
\end{figure*}

The computational performance of the different controllers and for the different number of subsystem cases is shown in Figures~\ref{fig:CompTimesI1}. The computation time represents the average time the numerical simulation takes from the initial sample time instant to that where all the plant states reach 0.02\% of their equilibrium value. This definition is chosen to correctly capture the behavior of all subsystems equally. Note that the IF-mpDiMPC approach solves simultaneous equations at each time step, making the overall computation time linearly proportional to the chosen simulation time (which is taken as 100 s with a 1 s time step in this study). It can be observed that all the controllers show an increasing trend in computation time with increase in the number of subsystems. In the case of IF-mpDiMPC, the computation time rises exponentially with the increase in the number of subsystems. Hence, solving IF-mpDiMPC for 4 and 5 subsystems proved to be computationally prohibitively expensive. This underscores the need for more efficient algorithms like IF-mpDiMPC-V2 when dealing with larger, more complex systems. The mean computation time for various control architectures is shown in Table~\ref{tab: MeanCompTims}.  The mean values of both IF-mpDiMPC and IF-mpDiMPC-V1.5 architectures are more than that of I-mpDiMPC for all the number of subsystem cases. However, the computation time of IF-mpDiMPC-V2 is less than that of I-mpDiMPC, again for all number of subsystem cases. Although the simultaneous solution approach of IF-mpDiMPC and IF-mpDiMPC-V1.5 is computationally expensive, for the case studies presented in the paper, it helps reduce communication load significantly compared to I-mpDiMPC as the information between the local controller is exchanged only once. Note that an increase in the number of iterations of I-mpDiMPC with increased communication load will also delay control input calculation, which is not accounted for in the numerical simulation study. The IF-mpDiMPC-V2 algorithm performs the best, not only in terms of the communication load reduction but also in the computation time reduction. 
\begin{table}
    \centering
    \caption{Average computation times (in seconds) for the different DiMPC control architectures and different numbers of subsystems.}
    \small
    \begin{tabular}{@{}lcccc@{}}
    \hline
    \textbf{Architecture} & \textbf{2} & \textbf{3} & \textbf{4} & \textbf{5}\\
    \hline
    DiMPC & 9.914 & 24.884 & 54.497 & 77.585  \\
    I-mpDiMPC & 0.074 & 0.515 & 3.254 & 12.777 \\
    IF-mpDiMPC & 0.354 & 167.218 & -- & -- \\
    IF-mpDiMPC-V1.5 & 0.617 & 3.174 & 12.374 & 67.993 \\
    IF-mpDiMPC-V2 & 0.205 & 0.511 & 0.758 & 2.936 \\
    \hline
    \end{tabular}
    \label{tab: MeanCompTims}
\end{table}

The maximum number of single sample time step intermediate iterations for the entire simulation time period for each experiment is recorded. Further, maximum and average values of all the maximum iteration numbers for DiMPC and I-mpDiMPC are determined and listed in Table~\ref{tab: maxAndAvgIter}. As can be seen, an increase in the number of subsystems results in an increase in both the maximum and average number of iterations. As the number of subsystems increases, obtaining a plantwide optimum through the DiMPC solution algorithm results in an increased number of intermediate iterations. The trend agrees with the results of Wang and Yang \cite{wang2022improved}, wherein the authors have compared the performance of DiMPC with their own improved iterative solution method.

To address the communication burden of different control architectures, the instances of data transfer between subsystems for all methods are analyzed. Figure~\ref{fig:ComuTimes} illustrates this comparison in both linear and logarithmic scales. The iteration-free methods (IF-mpDiMPC, IF-mpDiMPC-V1.5, and IF-mpDiMPC-V2) demonstrate significantly fewer instances of data transfer compared to the iterative methods (DiMPC and I-mpDiMPC), particularly as the number of subsystems increases. For iterative methods, the number of data transfers directly corresponds to the number of intermediate iterations required to converge, as reported in Table~\ref{tab: maxAndAvgIter}. In contrast, iteration-free methods require only a single instance of data transfer per time step (100 transfer instances for the entire simulation of 100 time steps), regardless of the number of subsystems. This substantial reduction in data transfer instances translates to decreased network traffic and lowered susceptibility to communication-related issues. While this analysis does not account for all potential sources of communication delay, it demonstrates how the proposed iteration-free methods significantly reduce the communication load, addressing a key motivation of this work.

\begin{table}
    \centering
     \caption{Maximal and average number of iterations under DiMPC (and mp-DiMPC).}
     \small
    \begin{tabular}{@{}cccc@{}}
    \hline
    \textbf{No. of Subsystems} & \textbf{Maximal Iteration} & \textbf{Average Iteration} \\
    \hline
     $2$ & $38$ & $19.99$  \\
    $3$ &  $62$ & $33.97$ \\
    $4$  &  $100$  &  $49.86$ \\
    $5$  & $100$ & $66.05$ \\
    % $6$  & $100$ & $82.77$ \\
    \hline
    \end{tabular}
    \label{tab: maxAndAvgIter}
\end{table}

\begin{figure*}[ht!]
  \centering
  \includegraphics[width=1\textwidth]{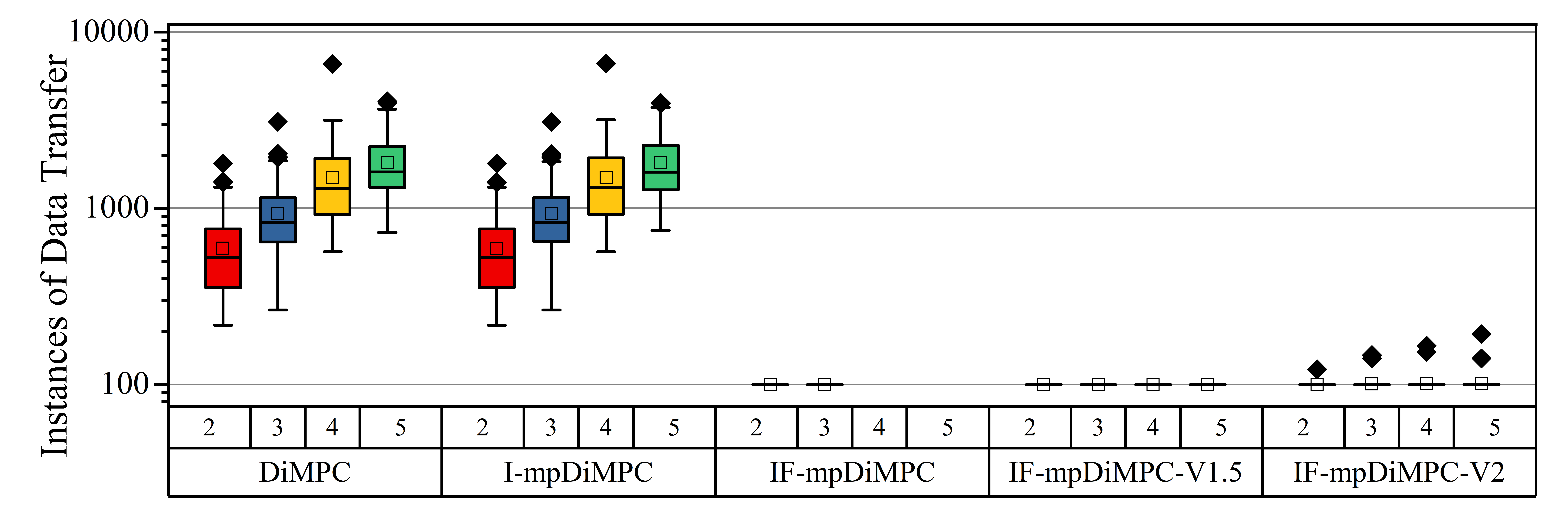}
  \caption{Number of instances of data transfer in log scale for the different DiMPC control architectures for the entire simulation of 100-time steps.}
  \label{fig:ComuTimes}
\end{figure*}

\section{Conclusions} \label{sec: conclusions}

Novel iteration-free DiMPC algorithms were developed based on multiparametric programming, IF-mpDiMPC, IF-mpDiMPC-V1.5, and IF-mpDiMPC-V2. Their communication and computational performances were studied and compared against their iterative counterparts. Randomly generated plants with different numbers of subsystems were used to perform numerical simulation studies of the closed-loop operation under the developed controllers. All the controllers were tuned to obtain the centralized-like output tracking performance for comparison purposes. In IF-mpDiMPC, the explicitly generated solutions of all the local controllers were solved simultaneously to obtain the control inputs at each sample time step. This method was improved in IF-mpDiMPC-V1.5 by performing feasibility checks on the explicitly generated solutions, in terms of constraint satisfaction, thereby reducing the set of equations that should be considered for simultaneous solutions online. This method was improved further in IF-mpDiMPC-V2 by only considering the previous sample time step solution and its nearest neighbors for the simultaneous solution online. IF-mpDiMPC and IF-mpDiMPC-V1.5 needed more time to solve and obtain the solution compared to the iterative I-mpDiMPC algorithm for all the number of subsystem cases as the time required to solve all the explicitly generated equations exceeded that of the iterative procedure convergence. IF-mpDiMPC-V2, on the other hand, performed the best in terms of the computation time among all the controllers and for all the number of subsystem cases due to the reduced set of equations for simultaneous solutions. Furthermore, the developed iteration-free controllers significantly reduced the communication burden, especially as system complexity increased, by requiring only a single exchange of information between local controllers at each sample time, in contrast to the multiple exchanges typical of iterative approaches. This streamlined data exchange improved the robustness of the control network in dealing with communication challenges.

A limitation of this work is that the computational cost of the developed IF-mpDiMPC approaches increases with the number of critical regions, which grows as the state dimension, prediction horizon, and control horizon increase. While IF-mpDiMPC-V2 partially mitigates this by checking only the current and neighboring critical regions, costs can still rise if the solution lies outside these areas. Additionally, this study has not yet explored worst-case scenarios, as the focus was on optimizing performance under nominal conditions. Future research could extend these methods to assess robustness and stability in such cases.

\bibliographystyle{elsarticle-num-names}
\bibliography{Manuscript_Arxiv_V1}

%%%%%%%%%% Merge with supplemental materials %%%%%%%%%%
\pagebreak
% \widetext
\begin{center}
\textbf{\large Supplemental Material: Iteration-Free Cooperative Distributed MPC through Multiparametric Programming}
\end{center}
%%%%%%%%%% Merge with supplemental materials %%%%%%%%%%
%%%%%%%%%% Prefix a "S" to all equations, figures, tables and reset the counter %%%%%%%%%%
\setcounter{equation}{0}
\setcounter{figure}{0}
\setcounter{table}{0}
\setcounter{page}{1}
\makeatletter
\renewcommand{\theequation}{S\arabic{equation}}
\renewcommand{\thefigure}{S\arabic{figure}}
\renewcommand{\bibnumfmt}[1]{[S#1]}
\renewcommand{\citenumfont}[1]{S#1}
%%%%%%%%%% Prefix a "S" to all equations, figures, tables and reset the counter %%%%%%%%%%
\section{Supporting Information}

This section contains additional figures that support the analysis presented in the main manuscript. All figures are labeled and described in detail below.

\begin{figure}[H]
  \centering
        \begin{subfigure}[b]{0.5\columnwidth}
            \includegraphics[width=\linewidth]{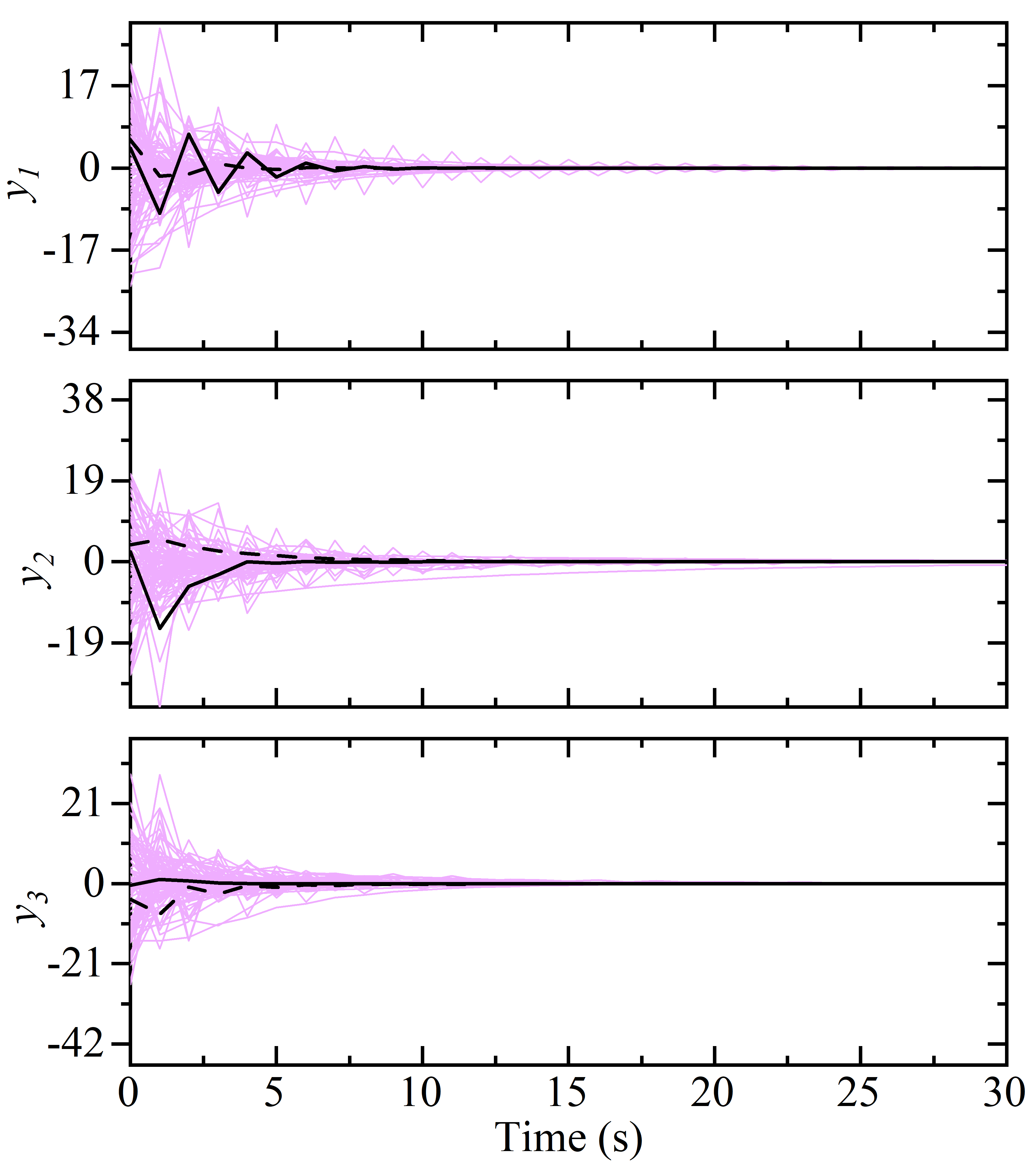}
            \caption{}
            \label{fig:3SubsOP}
        \end{subfigure}%
        \begin{subfigure}[b]{0.5\columnwidth}
            \includegraphics[width=\linewidth]{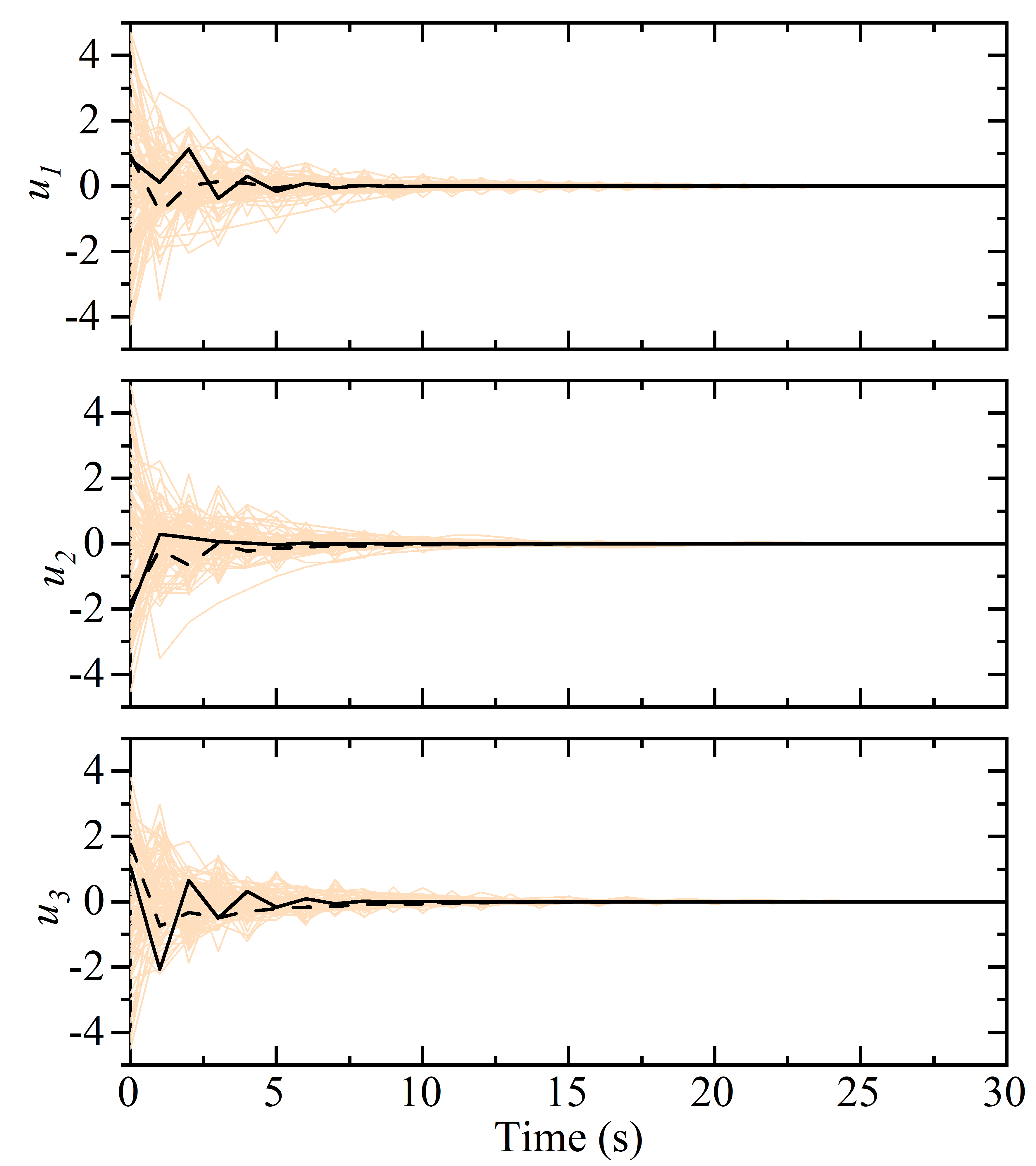}
            \caption{}
            \label{fig:3SubsIP}
        \end{subfigure}\\
  \caption{Controller performance for 3 subsystems. (a) Outputs of each subsystem. (b) Applied inputs to achieve the given outputs. The solid and dashed lines represent two random experimental runs in the output and input plots for visualization purposes.}
  \label{fig:3SubsPerformances}
\end{figure}

\begin{figure}[H]
  \centering
        \begin{subfigure}[b]{0.5\columnwidth}
            \includegraphics[width=\linewidth]{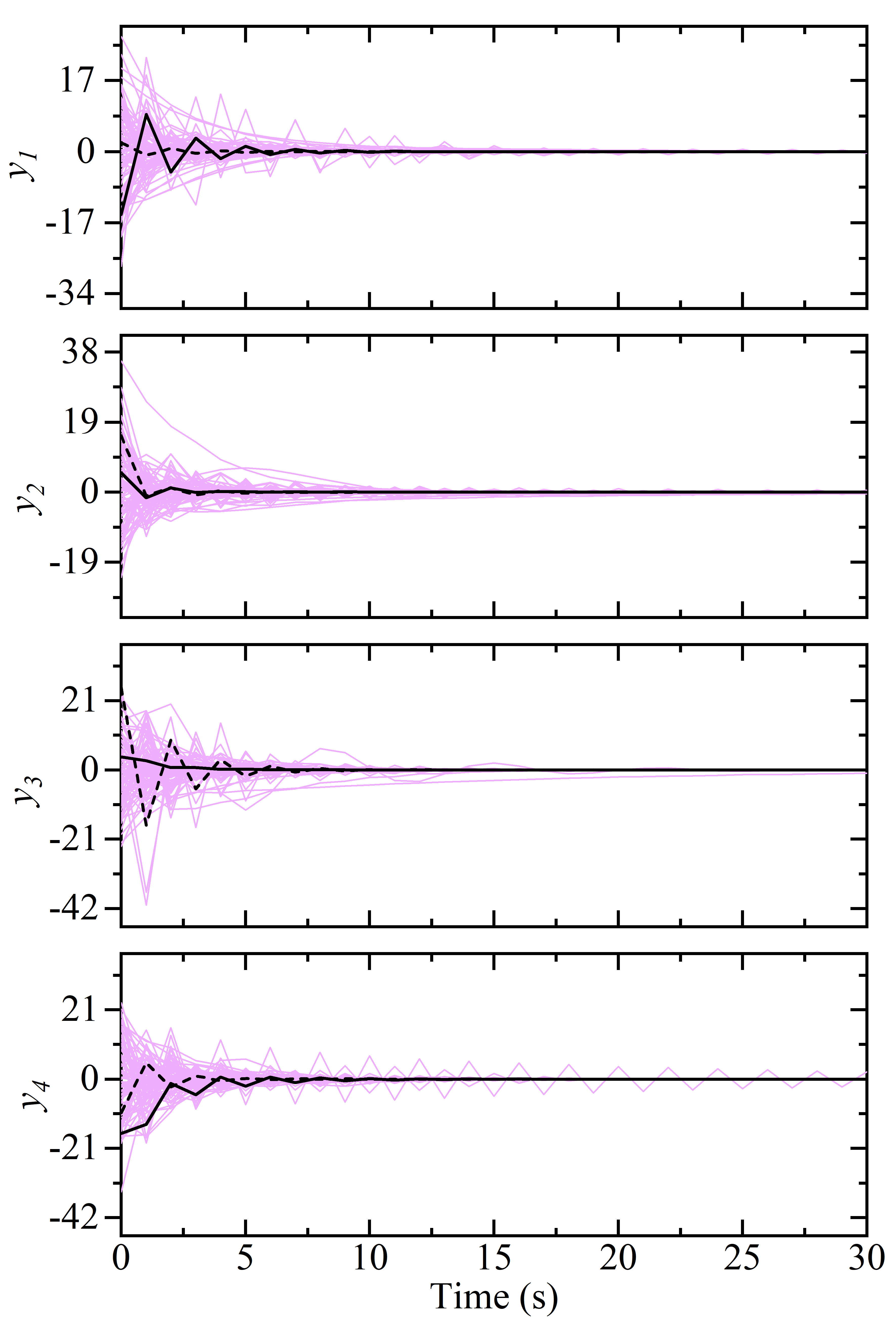}
            \caption{}
            \label{fig:4SubsOP}
        \end{subfigure}%
        \begin{subfigure}[b]{0.5\columnwidth}
            \includegraphics[width=\linewidth]{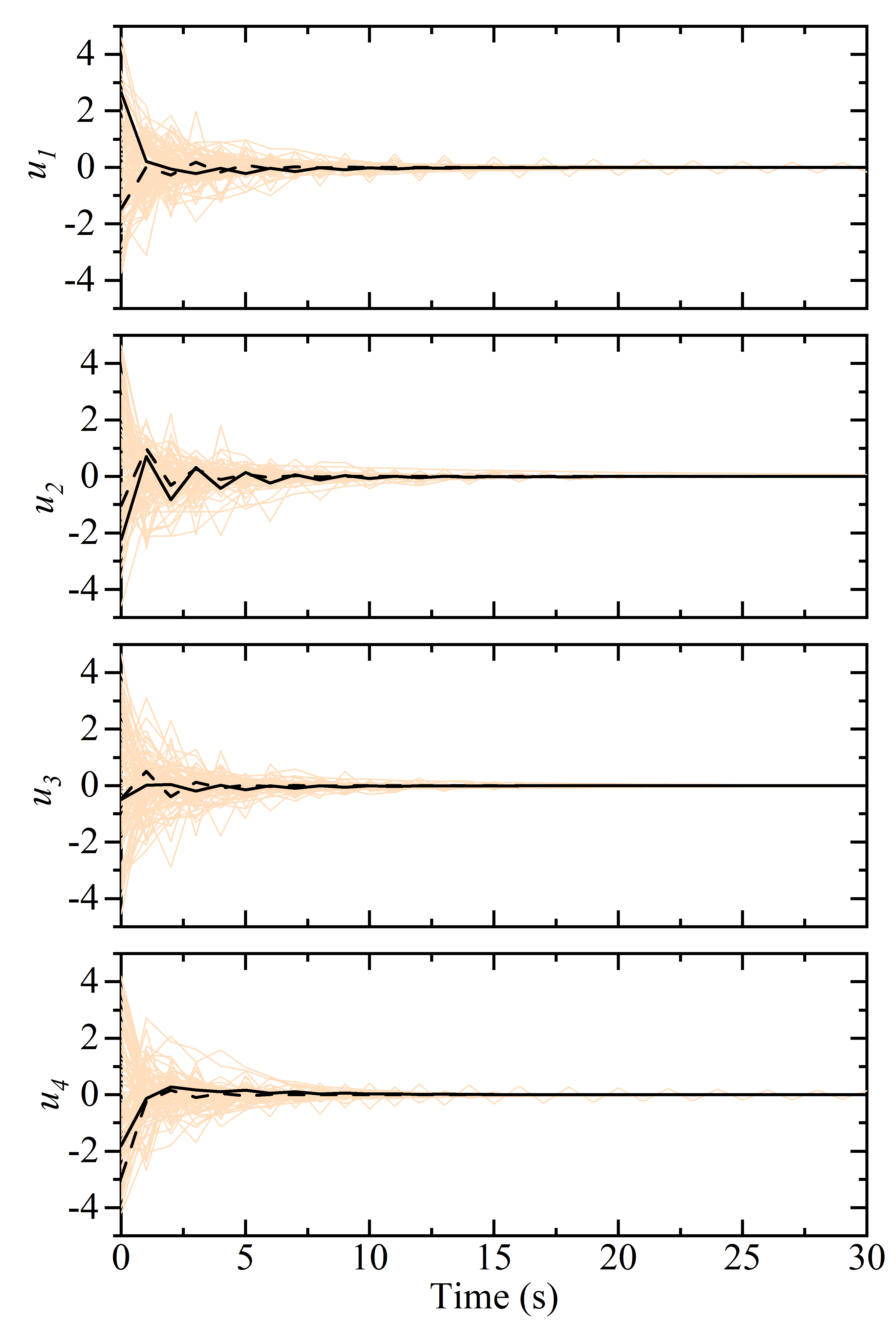}
            \caption{}
            \label{fig:4SubsIP}
        \end{subfigure}\\
  \caption{Controller performance for 4 subsystems. (a) Outputs of each subsystem. (b) Applied inputs to achieve the given outputs. The solid and dashed lines represent two random experimental runs in the output and input plots for visualization purposes.}
  \label{fig:4SubsPerformances}
\end{figure}

\begin{figure}[H]
  \centering
        \begin{subfigure}[b]{0.5\columnwidth}
            \includegraphics[width=\linewidth]{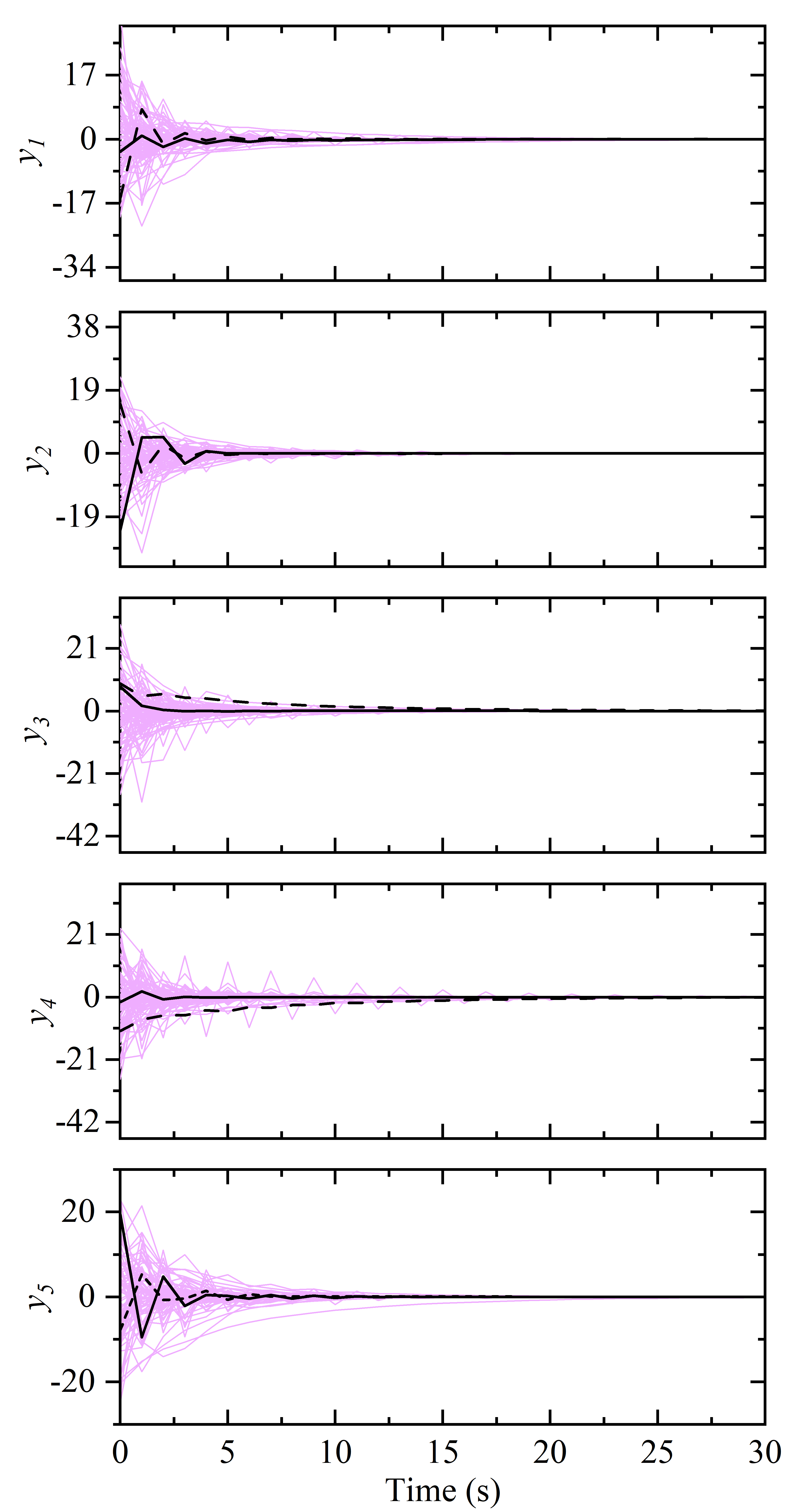}
            \caption{}
            \label{fig: 5SubsOP}
        \end{subfigure}%
        \begin{subfigure}[b]{0.5\columnwidth}
            \includegraphics[width=\linewidth]{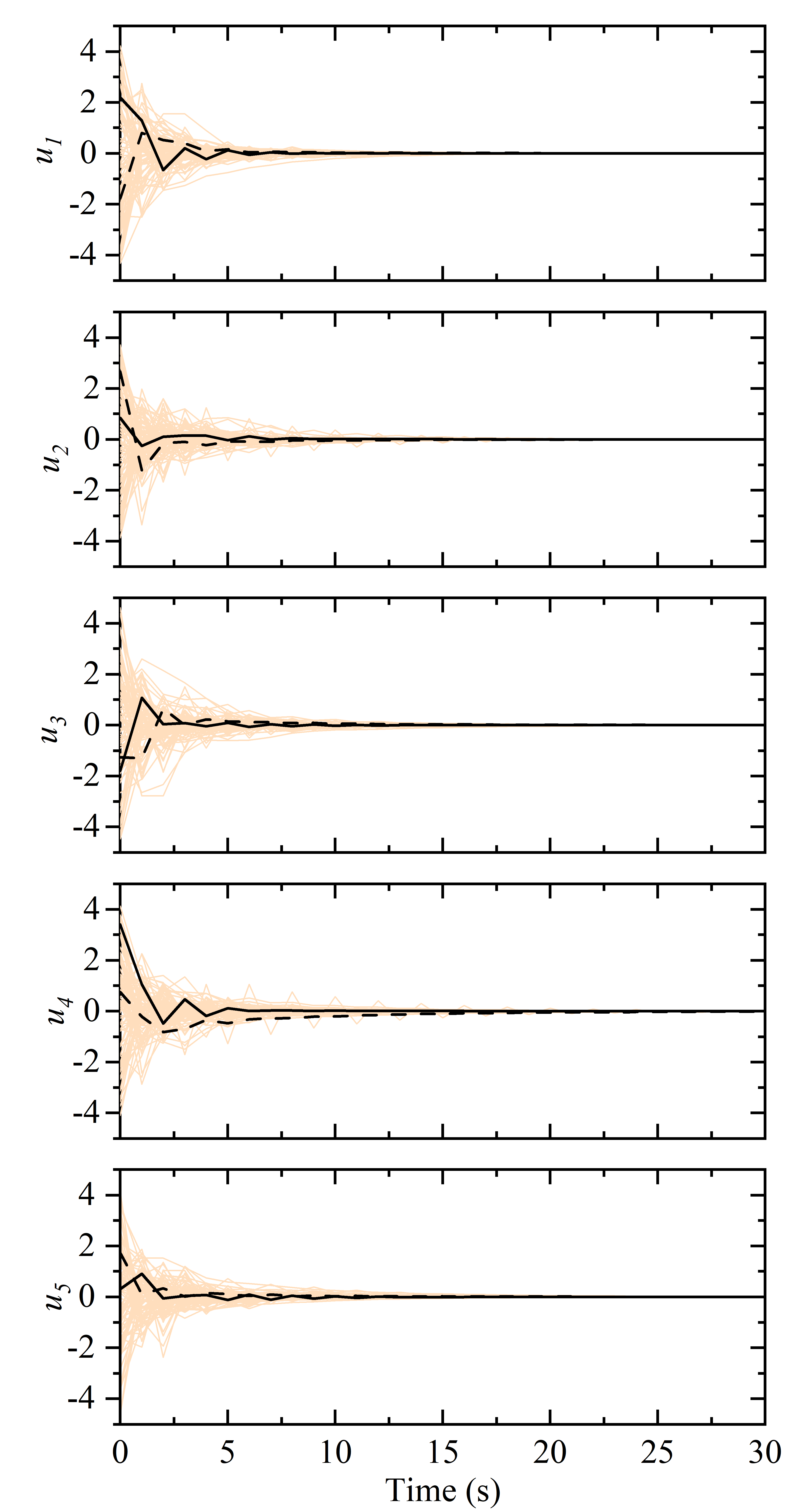}
            \caption{}
            \label{fig: 5SubsIP}
        \end{subfigure}\\
  \caption{Controller performance for 5 subsystems.(a) Outputs of each subsystem. (b) Applied inputs to achieve the given outputs. The solid and dashed lines represent two random experimental runs in the output and input plots for visualization purposes.}
  \label{fig:5SubsPerformances}
\end{figure}

\begin{figure}[H]
  \centering
  \includegraphics[width=0.95\textwidth]{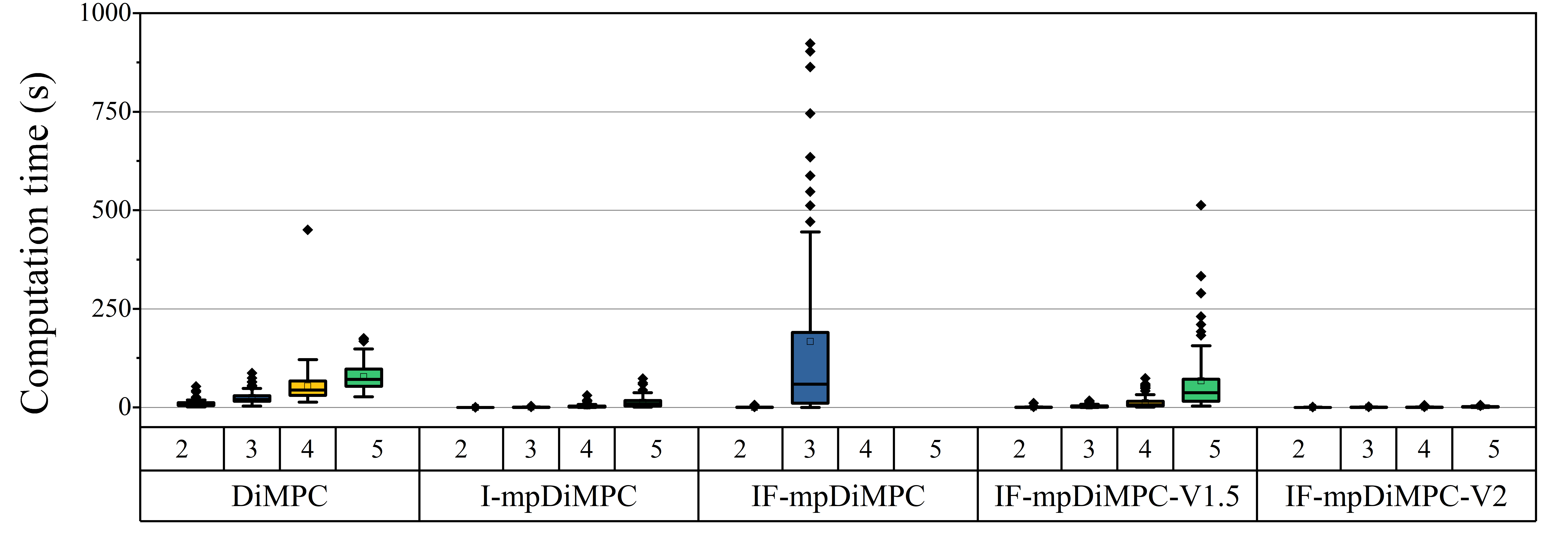}
  \caption{Computation times in linear scale for the different DiMPC control architectures.}
  \label{fig:CompTimesI1}
\end{figure}

\begin{figure}[H]
  \centering
        \begin{subfigure}[b]{0.33\textwidth}
            \includegraphics[width=\linewidth]{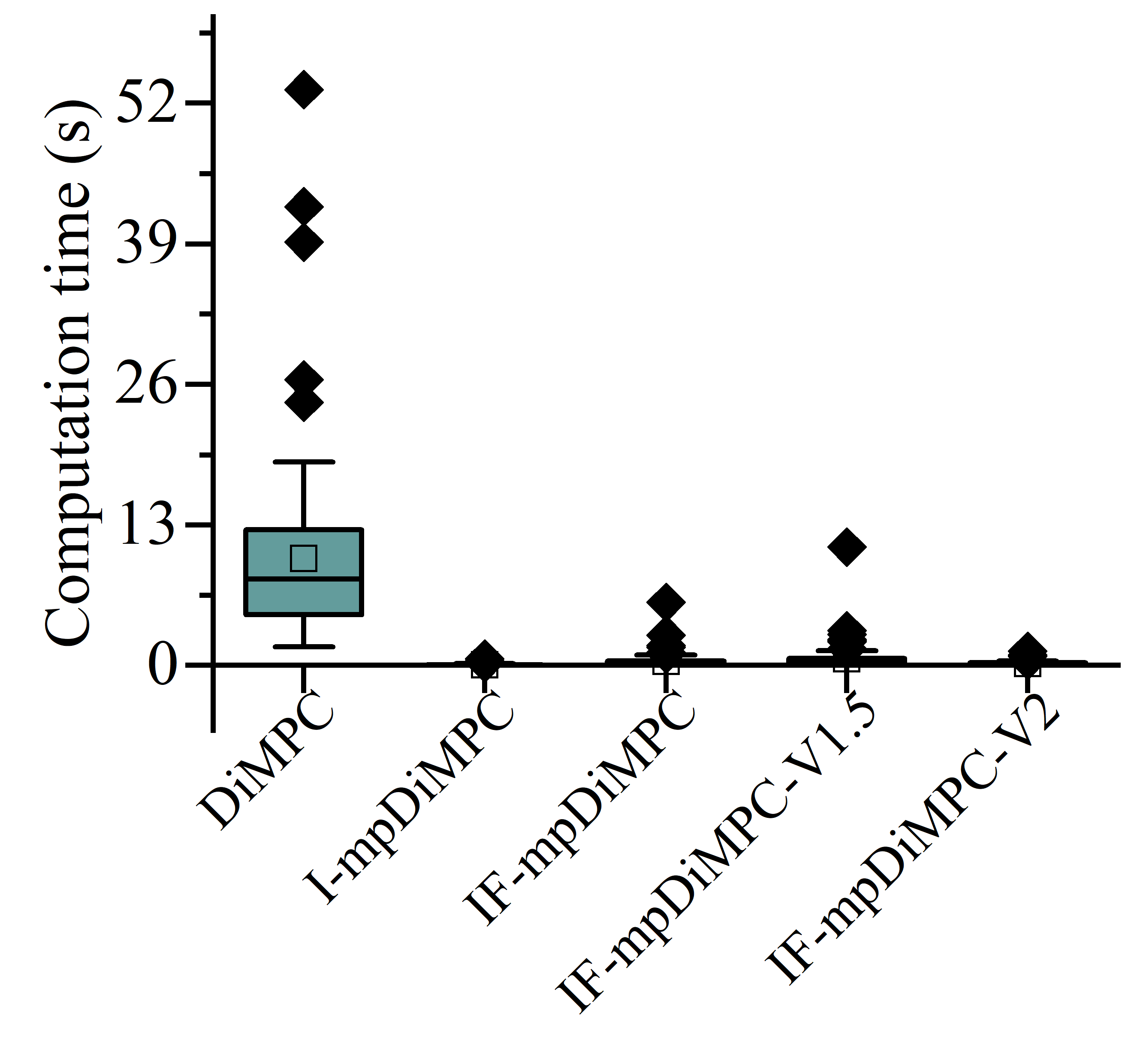}
            \caption{}
            \label{fig: 2SubsLinear}
        \end{subfigure}%
        \begin{subfigure}[b]{0.33\textwidth}
            \includegraphics[width=\linewidth]{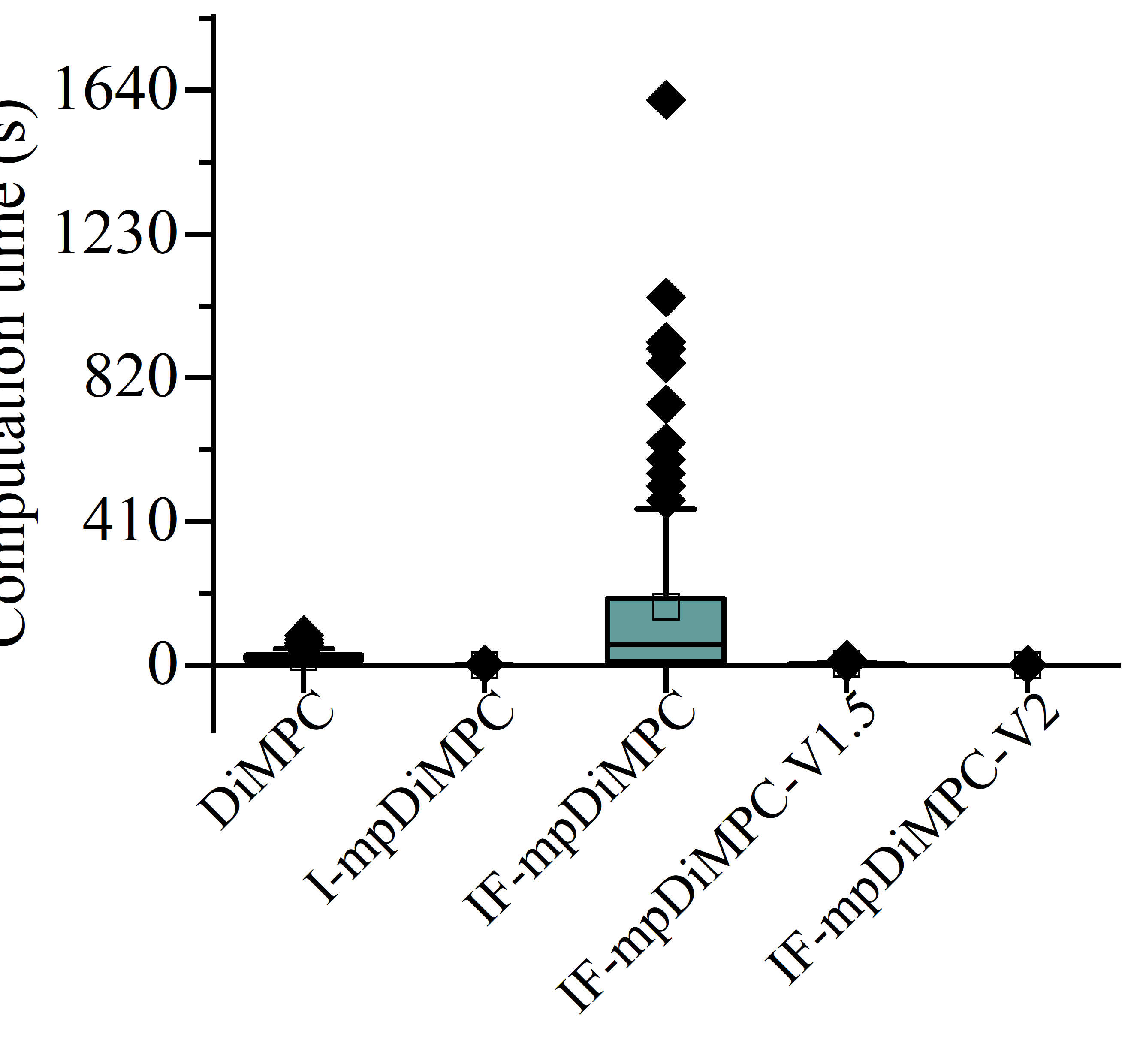}
            \caption{}
            \label{fig: 3SubsLinear}
        \end{subfigure}%
        \begin{subfigure}[b]{0.35\textwidth}
            \includegraphics[width=\linewidth]{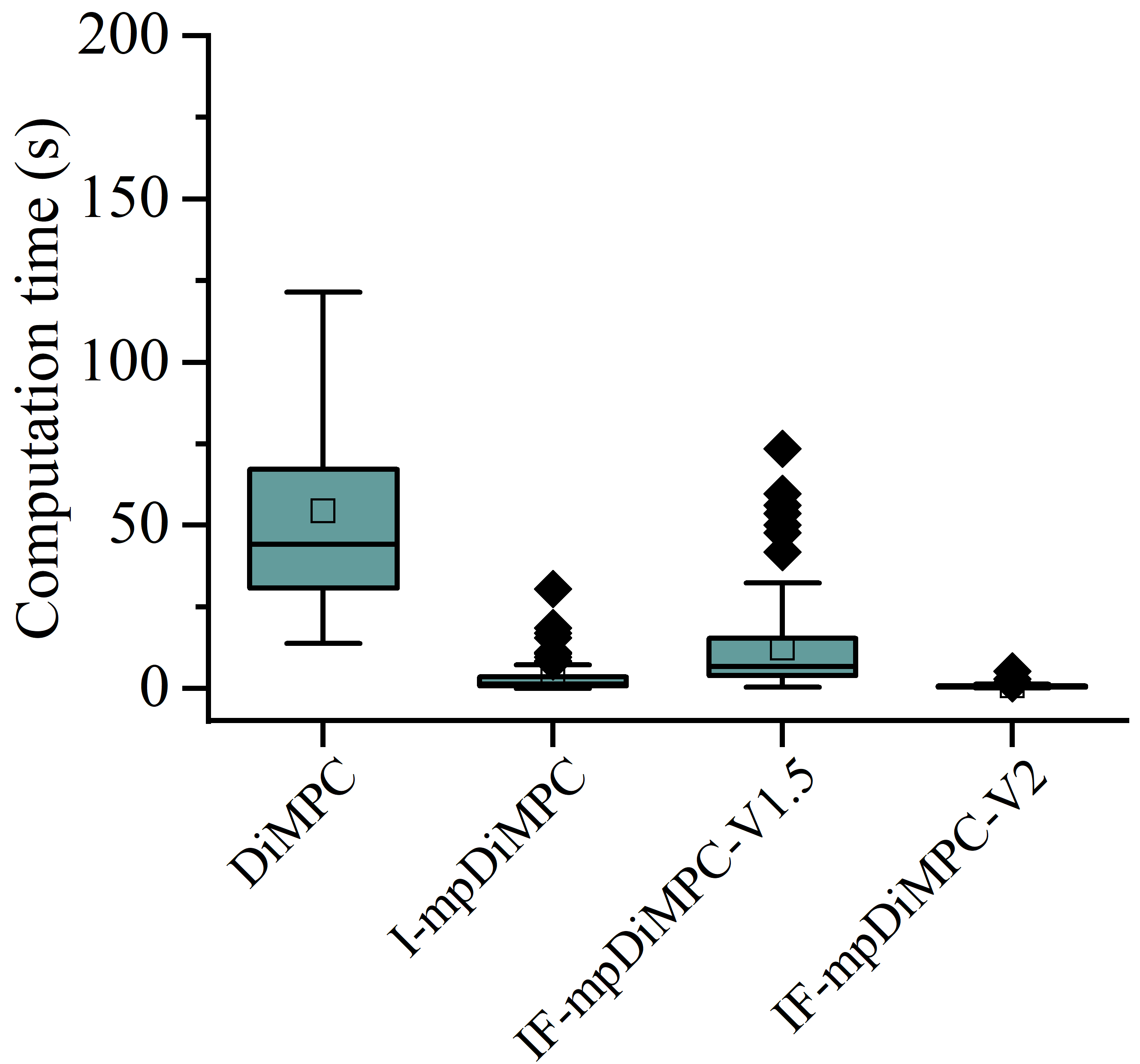}
            \caption{}
            \label{fig: 4SubsLinear}
        \end{subfigure}\\
        \begin{subfigure}[b]{0.33\textwidth}
            \includegraphics[width=\linewidth]{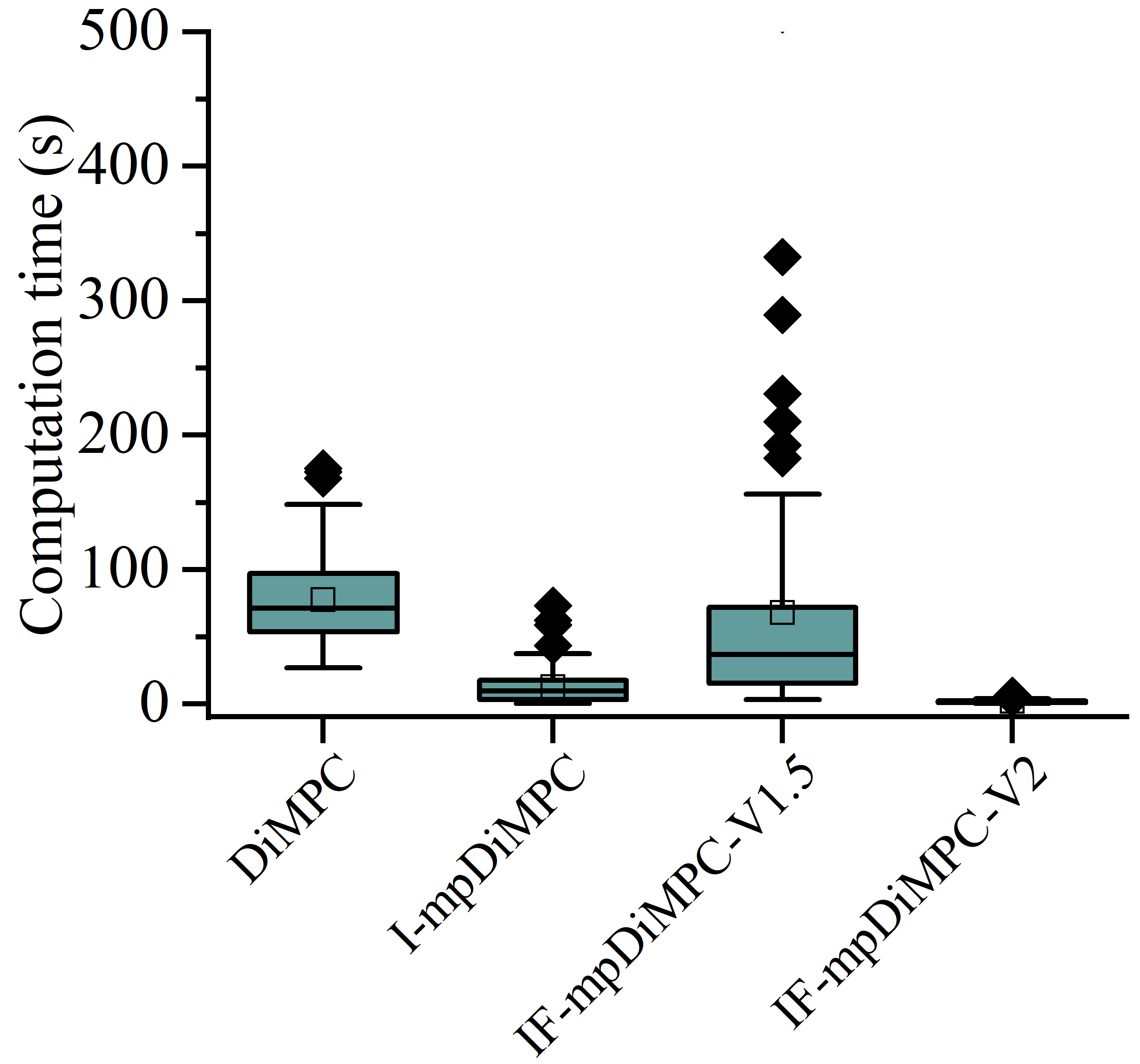}
            \caption{}
            \label{fig: 5SubsLinear}
        \end{subfigure}%
    \caption{Illustration 1 (Part 1): Computation times for 2, 3, 4, and 5 subsystem case are plotted in linear scale in (a) - (d), respectively, for the different DiMPC control architectures.}
  \label{fig:CompTimesLine}
\end{figure}
            
\begin{figure}[H]
  \centering
        \begin{subfigure}[b]{0.33\textwidth}
            \includegraphics[width=\linewidth]{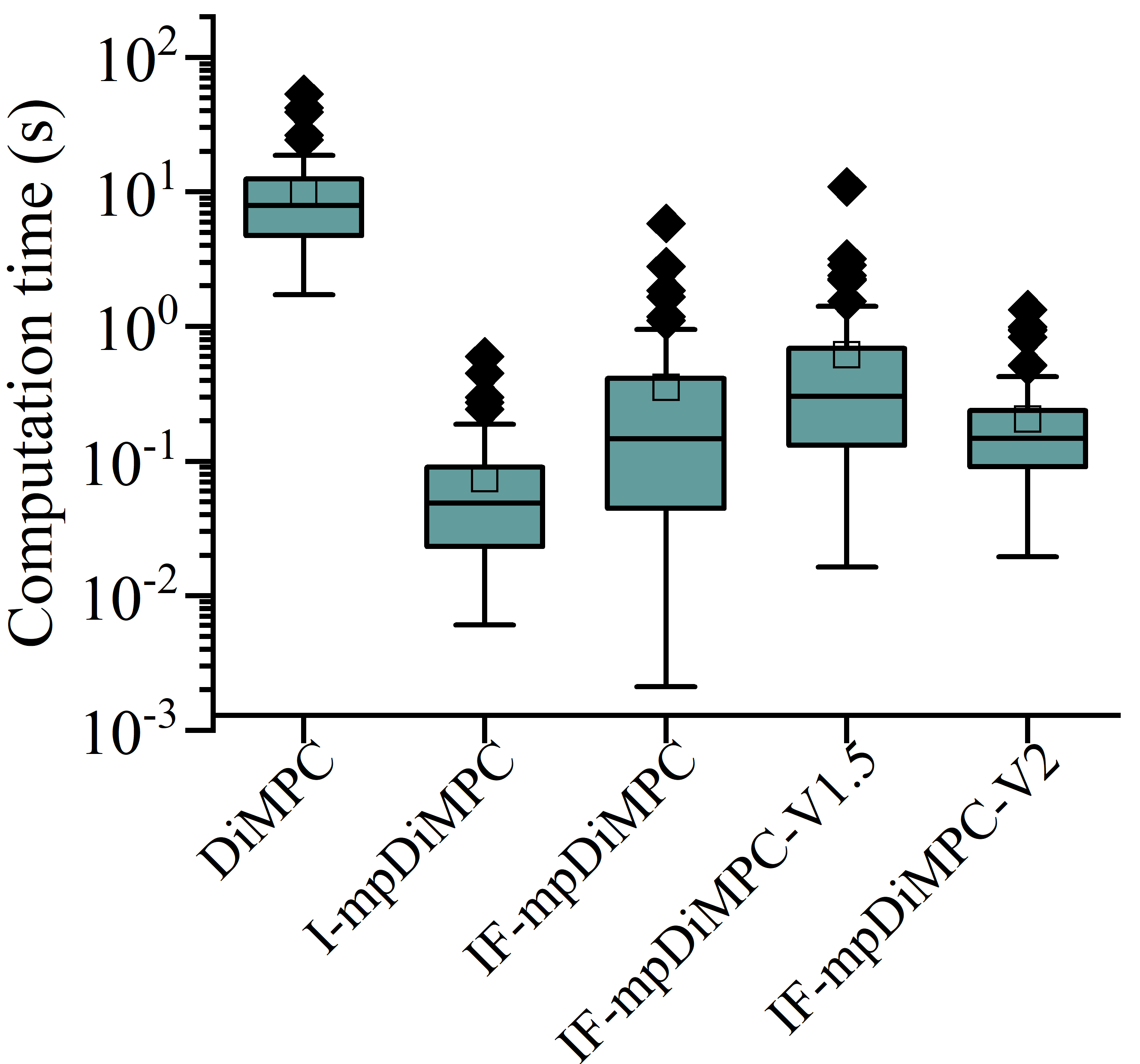}
            \caption{}
            \label{fig: 2SubsLog}
        \end{subfigure}%
        \begin{subfigure}[b]{0.33\textwidth}
            \includegraphics[width=\linewidth]{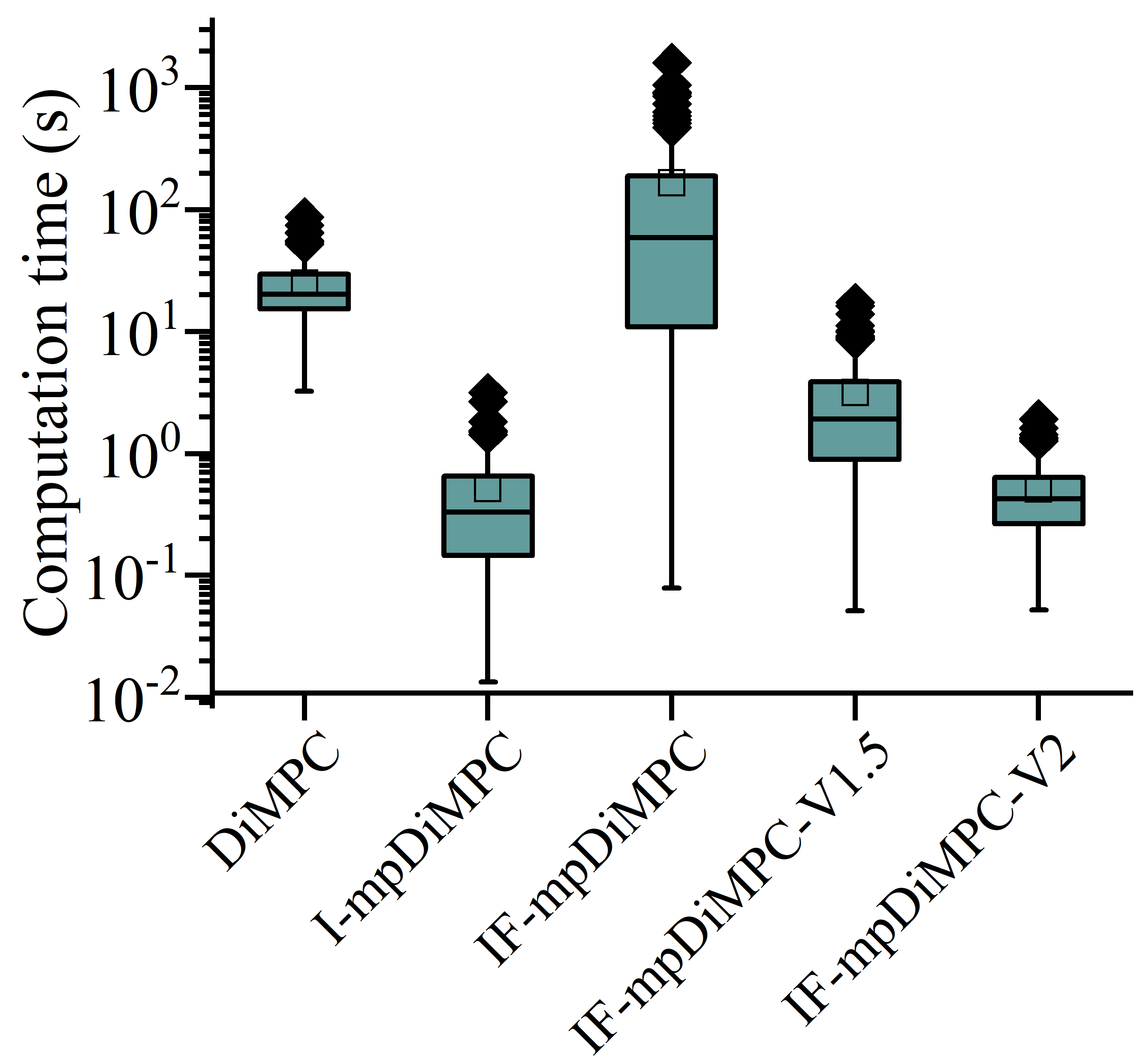}
            \caption{}
            \label{fig: 3SubsLog}
        \end{subfigure}%
        \begin{subfigure}[b]{0.33\textwidth}
            \includegraphics[width=\linewidth]{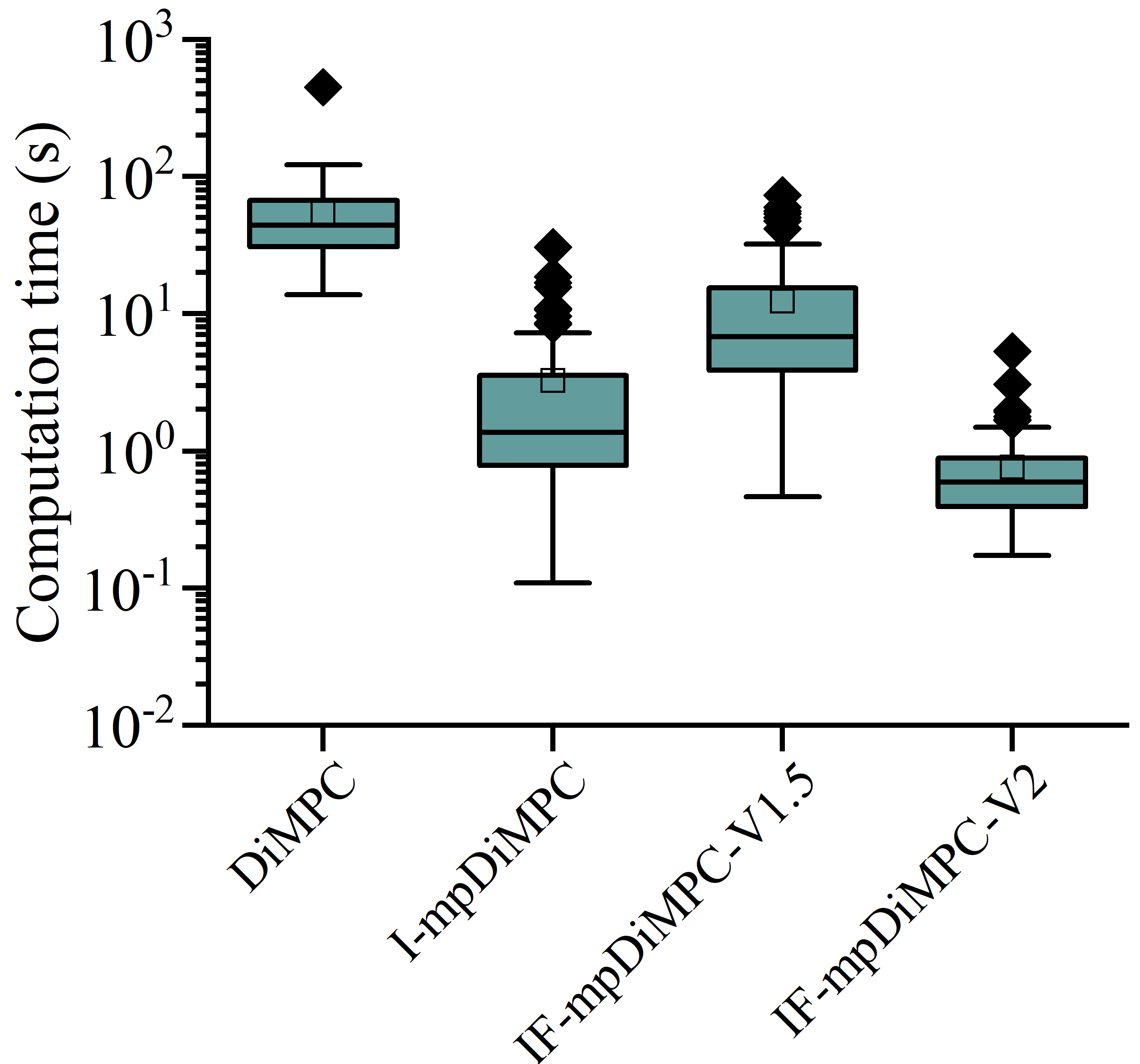}
            \caption{}
            \label{fig: 4SubsLog}
        \end{subfigure}\\
        \begin{subfigure}[b]{0.33\textwidth}
            \includegraphics[width=\linewidth]{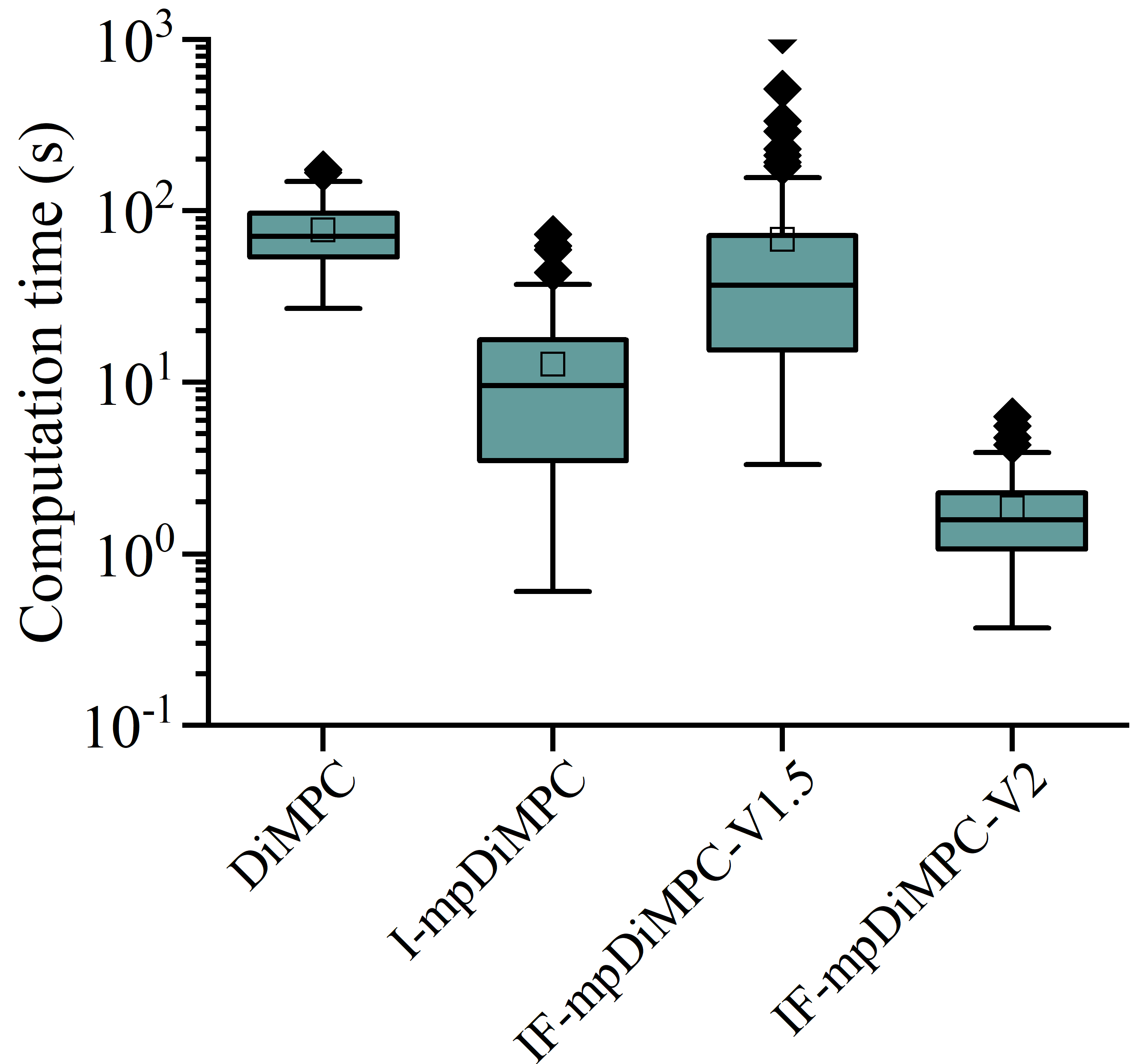}
            \caption{}
            \label{fig: 5SubsLog}
        \end{subfigure}%
  \caption{Illustration 1 (Part 2): Computation times for 2, 3, 4, and 5 subsystem case are plotted in log scale in (a) - (d), respectively, for the different DiMPC control architectures.}
  \label{fig:CompTimesLog}
\end{figure}

\begin{figure}[H]
  \centering
  \includegraphics[width=0.95\textwidth]{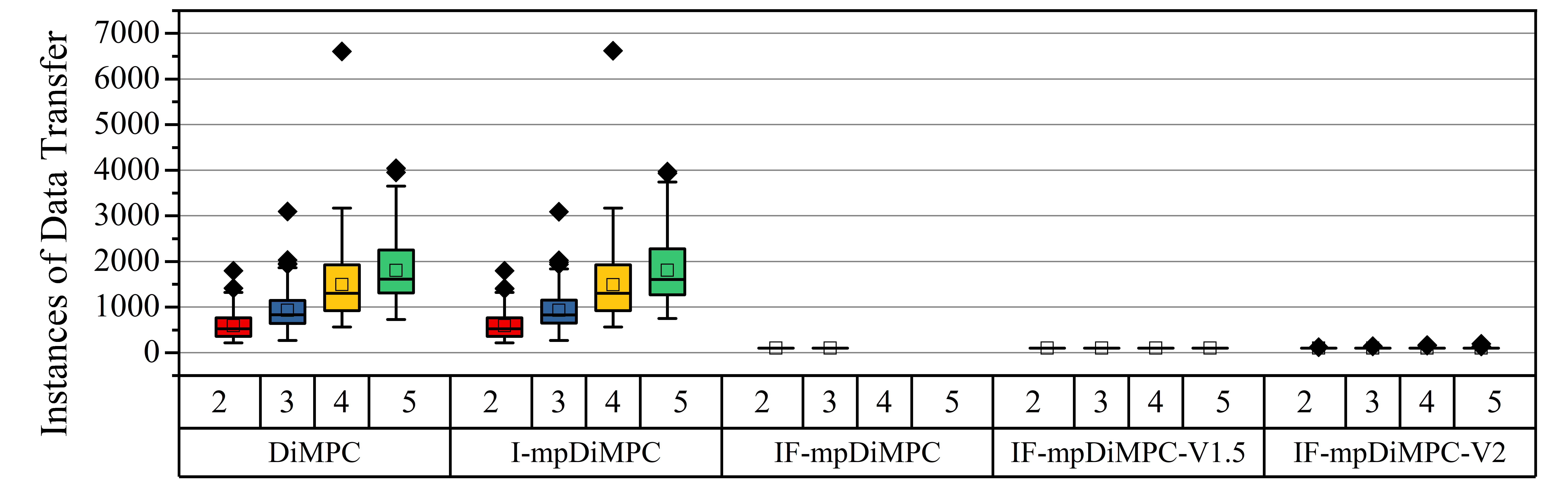}
  \caption{Number of instances of data transfer in linear scale for the different DiMPC control architectures for the entire simulation of 100 time steps.}
  \label{fig:ComuTimes}
\end{figure}

\end{document}